\documentclass{jfm}
\usepackage{graphicx}
\usepackage{epstopdf, epsfig}
\usepackage{amsmath}
\usepackage{multicol}
\usepackage{empheq}
\usepackage{float}
\usepackage{xcolor}
\usepackage{optidef}
\usepackage{lineno}
\usepackage{optidef}
\usepackage[ruled,vlined]{algorithm2e}

\shorttitle{\textit{A-priori} sparsification of Galerkin based reduced order models}
\shortauthor{R. Rubini, D. Lasagna and A. Da Ronch}

\title{A-priori sparsification of Galerkin-based reduced order models}

\author{Riccardo Rubini\aff{1},
        Davide Lasagna\aff{1}\corresp{\email{davide.lasagna@soton.ac.uk}}
  \and Andrea Da Ronch\aff{1}}

\affiliation{\aff{1}Faculty of Engineering and Physical Sciences, University of Southampton, SO17 1BJ Southampton, UK}

\begin{document}

\maketitle

\begin{abstract}
A methodology to generate sparse Galerkin models of chaotic/unsteady fluid flows containing a minimal number of active triadic interactions is proposed. The key idea is to find an appropriate set of basis functions for the projection representing elementary flow structures that interact minimally one with the other and thus result in a triadic interaction coefficient tensor with sparse structure. Interpretable and computationally efficient Galerkin models can be thus obtained, since a reduced number of triadic interactions needs to be computed to evaluate the right hand side of the model. To find the basis functions, a subspace rotation technique is used, whereby a set of Proper Orthogonal Decomposition (POD) modes is rotated into a POD subspace of larger dimension using coordinates associated to low-energy dissipative scales to alter energy paths and the structure of the triadic interaction coefficient tensor. This rotation is obtained as the solution of a non-convex optimisation problem that maximises the energy captured by the new basis, promotes sparsity and ensures long-term temporal stability of the sparse Galerkin system. We demonstrate the approach on two-dimensional lid-driven cavity flow at $Re = 2 \times 10^4$ where the motion is chaotic. We show that the procedure generates Galerkin models with a reduced set of active triadic interactions, distributed in modal space according to established knowledge of scale interactions in two-dimensional flows. This property, however, is only observed if long-term temporal stability is explicitly included in the formulation, indicating that a dynamical constraint is necessary to obtain a physically consistent sparsification. 
\end{abstract}


\section{Introduction}\label{sec:intro}
%
From a physical standpoint, turbulence is a multi-scale phenomenon exhibiting a wide hierarchy of spatial and temporal scales. This property, coupled with the intrinsic nonlinearity of the governing equations, poses considerable difficulties to its modelling and analysis. One of the major challenges to obtain a satisfactory mathematical description of this phenomenon arises from the fact that the dynamics of flow structures at a particular length or time scale cannot be examined in isolation without also considering at the same time the whole hierarchy of complementary scales \citep{domaradzki1994energy}. In fact, nonlinear interactions between triads of scales play a fundamental role as they are are the main driver of energy transfer between coherent structures \citep{pope2001turbulent, moffatt2014note}. In turn, the organisation of triadic interactions has a direct influence on the physics of a number of flow phenomena, such as direct and inverse energy cascades \citep{kolmogorov1991local} or transition to turbulence \citep{craik1971non, rempfer1994dynamics, rempfer1994evolution} in different flow configurations \citep{schmidt2020bispectral}. Overall, this property makes the development of computationally efficient and physically interpretable reduced-order dynamical models a challenging task.


Historically, the study of triadic interactions has been conducted by first employing an appropriate decomposition technique to educe coherent structures from the turbulent motion and then characterising the intensity of the inter-modal couplings through the perspective of the resulting Galerkin model \citep{noack2008finite}. For homogeneous isotropic turbulence, Fourier modes provide an optimal representation \citep{brasseur1994interscale,laval1999nonlocality}, but for flow in complex geometries, modes identified from data with Proper Orthogonal Decomposition (POD) \citep{couplet2003intermodal,rempfer1994dynamics} or with frequency-domain decomposition methods  \citep{symon2020energy,schmid2010dynamic,towne2018spectral} have been often adopted. 
One of the key findings of such studies is that energy transfers are not uniformly distributed in modal space. In fact, not all interactions have the same importance and energy flows along preferential directions. Specifically, there is evidence suggesting that the nonlinear interaction pattern among coherent structures is often sparse. In other words, the evolution of structures at a certain length scale depends predominantly upon a subset of all other structures \citep{kraichnan1971inertial,ohkitani1990nonlocality, brasseur1994interscale} and the influence of interactions with the complementary set of structures can be generally neglected with minor global effects. This behaviour has been observed for a multiplicity of different flows, ranging from bluff body wakes \citep{jin2021energy} to transitional boundary layers \citep{rempfer1994dynamics,rempfer1994evolution} and separated flows \citep{couplet2003intermodal}.

However, these studies have typically focused on the interaction between modes educed using a flow decomposition technique that does not necessarily capture faithfully or optimally the aforementioned physics of scale interactions, since no information regarding nonlinear mechanisms that may produce such interactions is utilised in the decomposition. Hence, rationalising scale interactions and identifying relevant physical mechanisms may be challenging, especially in large models. Research into these aspects has only recently started \citep{schmidt2020bispectral}.
In addition, when classical modal decompositions are employed for model reduction, the quadratic coefficient tensor of the resulting Galerkin model is generally dense \citep{rubini2020l1} implying that computational costs to evaluate the nonlinearity grow as the third power of the number of modes, quickly becoming intractable for high-Reynolds-number configurations \citep{cordier2010calibration}. 

To address this challenge, we recently utilised $l_1$-based regression methods \citep{brunton2016discovering, loiseau2018constrained} to extract sparsity patterns in the inter-modal energy transfers in large Galerkin models of multi-scale flows, to construct sparse, computationally efficient and interpretable models \citep{rubini2020l1}. The technique involves an \textit{a posteriori} evaluation of energy interactions between a pre-determined set of modes, pruning weak interactions (model coefficients) that do not contribute significantly to the overall dynamics. 
However, a rigorous connection between the modal structures and the projection model is necessarily lost, since the procedure involves tuning model coefficients without modifying the basis functions. Hence, the analysis of energy paths in the model is not necessarily consistent with the spatial structure of the basis functions. In addition, it was demonstrated that the organisation and sparsity of energy interactions is not invariant with respect to a change of the basis functions utilised for the analysis of the energy budget. This suggested that it might be possible to generate a set of modal structures producing a sparse Galerkin model, without requiring further tuning.

In this paper, we attempt to bridge this gap. We propose an \emph{a priori} sparsification method for Galerkin models where 
we seek a set of modal structures that capture energy transfer mechanisms efficiently, i.e. where the quadratic coefficient tensor regulating triadic interactions is as sparse as possible. The mathematical approach is inspired by the subspace rotation technique of \citet{balajewicz2013low, balajewicz2016minimal}, where a small rotation of the POD subspace was sought to absorb the unresolved dissipative scales into the basis functions set and stabilise the long-term behaviour without the need for empirical eddy-viscosity terms. Here, the key idea is to seek a small rotation of the original POD subspace to alter and sparsify energy transfer paths, without the need for an \emph{a posteriori} sparsification. In practice, the rotation is found by solving a constrained optimisation problem, minimising the loss of energy optimality subject to a constraint on the sparsity of the quadratic coefficient tensor. 


The manuscript is organised as follows. Section \ref{sec:methodology} outlines the general methodology to generate reduced order models by Galerkin projection and how energy interactions between modes defining such models can be analysed. Subsequently, the subspace rotation technique to generate sparse Galerkin models is outlined. In section \ref{sec:results}, we demonstrate this approach on two-dimensional lid-driven cavity flow at Reynolds number $Re = 2 \times 10^4$, where dynamics is chaotic \citep{auteri2002numerical} and inter-modal energy transfers are highly scattered in modal space.

\section{Methodology}\label{sec:methodology}

\subsection{Galerkin-based models and energy analysis}
We consider the space of square integrable solenoidal velocity vector fields defined over a spatial domain $\Omega$, endowed by the standard inner product
\begin{equation}\label{eq:inner_product}
    (\mathbf{u},\mathbf{v}) \coloneqq \int_{\Omega} \mathbf{u}\cdot\mathbf{v} \mathrm{d}\Omega, 
\end{equation}
where $\mathbf{u}, \mathbf{v}$ are two element of such space. 
The resulting $\mathcal{L}^2(\Omega)$ norm is denoted as $||\mathbf{u}||_2 = \sqrt{(\mathbf{u}, \mathbf{u})}$. Using the time averaged velocity field $\bar{\mathbf{u}}(\mathbf{x})$ as a base flow and denoting by $\mathbf{u}^\prime(t, \mathbf{x}) = \mathbf{u}(t, \mathbf{x}) - \bar{\mathbf{u}}(\mathbf{x})$ the velocity fluctuation, the $N$-dimensional Galerkin ansatz 
\begin{equation}\label{eq:modal_expansion}
	\mathbf{u}(t,\mathbf{x})  = \bar{\mathbf{u}}(\mathbf{x}) + \mathbf{u}^\prime(t, \mathbf{x}) \simeq \bar{\mathbf{u}}(\mathbf{x}) + \sum_{i = 1}^{N} a_i(t){\boldsymbol \phi}_i(\mathbf{x})
\end{equation}
is introduced to describe the space-time velocity field, where $a_i(t)$ and $\boldsymbol \phi_i(\mathbf{x})$, $i=1,\ldots N$ are the temporal and global spatial modes, respectively. We assume throughout that the spatial modes form an orthonormal set and satisfy the boundary conditions on $\Omega$.
Reduced order models are then derived by projecting the Navier-Stokes equations for incompressible flows onto the subspace defined by the spatial modes \citep{rowley2017model} using the inner product defined in \eqref{eq:inner_product}. 

Restricting the analysis to configurations where the boundaries are either no-slip walls or periodic, the pressure term arising from the projection vanishes identically \citep{schlegel_noack_2015}. Since the spatial modes define a solenoidal field, 
the system of coupled nonlinear differential equations (ODEs)
\begin{equation}\label{eq:galerkin}
	 \dot{a}_i(t) = \mathsf{C}_i + \sum_{j=1}^N \mathsf{L}_{ij}a_{j}(t) + \sum_{j=1}^N \sum_{k=1}^N \mathsf{Q}_{ijk}a_{j}(t)a_{k}(t), \quad \quad i=1\ldots,N
\end{equation}
is obtained, defining the temporal evolution of the coefficients $a_{i}(t)$ and where the tensors $\mathsfbi{C} \in \Re^{N}$, $\mathsfbi{L}  \in \Re^{N\times N}$ and $\mathsfbi{Q}  \in \Re^{N\times N\times N}$ are defined by suitable inner products involving the spatial modes. For instance, the entries of the quadratic coefficient tensor are
\begin{equation}
\mathsf{Q}_{ijk} = (\mathbf{ \boldsymbol{\phi}}_i, \mathbf{\boldsymbol{\phi}}_j \cdot \nabla \mathbf{\boldsymbol{\phi}}_k),
\label{eq:definition_qijk}
\end{equation}
while expressions for the constant and linear terms can be found, e.g., in \citet{noack2011reduced}. 

The expansion (\ref{eq:modal_expansion}) provides a suitable foundation to examine interactions between coherent structures in complex geometries. Here, we follow established approaches \citep{rempfer1994evolution}) and analyse such interactions by introducing the modal energies \mbox{$e_i(t) = \frac{1}{2} a_i(t)a_i(t)$, $i=1\ldots,N$}. For an expansion consisting of $N$ modes, the domain integral of the kinetic energy of velocity fluctuation expressed by the ansatz \eqref{eq:modal_expansion} is then given by
\begin{equation}\label{eq:energy_cont}
   E(t) =  \sum_{i=1}^{N} e_{i}(t)
\end{equation}
The instantaneous rate of change of the modal energies is given by 
\begin{equation}\label{eq:energy_POD}
    \dot{e}_i(t) = \mathsf{C}_i a_i(t) + \sum_{j = 1}^{N}\mathsf{L}_{ij} a_{i}(t) a_{j}(t) + \sum_{j = 1}^{N}\sum_{k = 1}^{N} \mathsf{Q}_{ijk}a_{i}(t)a_{j}(t)a_{k}(t), \quad \quad i=1,\ldots,N,
\end{equation}
obtained by multiplying (\ref{eq:galerkin}) by $a_i(t)$. The right hand side of equation (\ref{eq:energy_POD}) is composed of three terms describing energy transfers between the hierarchy of modes \citep{noack2008finite, noack2011reduced}. The first two describe variations of energy due to production/dissipation arising from interactions with the mean flow and from viscous effects. The third term defines variations of energy arising from inviscid nonlinear interactions between triads of modes. Additional insight can be gained by taking the temporal average of equation \eqref{eq:energy_POD}, i.e. by examining the time-averaged energy budget of  system \eqref{eq:galerkin}. Assuming that the temporal coefficients have zero mean, we obtain
\begin{equation}\label{eq:energy_avg}
    \sum_j^N  \mathsf{L}_{ij}\overline{a_i a_j} + \sum_{j=1}^N \sum_{k=1}^N \mathsf{Q}_{ijk}\overline{a_ia_ja_k} \simeq 0, \quad i = 1, \ldots, N,
\end{equation} 
where the summation over $j$ in the linear term accounts for the case where temporal coefficients are not uncorrelated in time. 
Equation \eqref{eq:energy_avg} shows that, on average, the energy balance is regulated by the production/dissipation described in the linear term and by the nonlinear energy transfer rate described by the quadratic term. As explained in \citet{balajewicz2013low}, the residual of (\ref{eq:energy_avg}) only vanishes when $N \to \infty$ and it is generally expected to be positive for finite-dimensional Galerkin models.

Following \cite{rempfer1994dynamics}, to better visualise the relative importance of the triadic interactions in an average we introduce the tensor $\mathsfbi{N} \in \Re^{N\times N\times N}$ with entries
\begin{equation}\label{eq:Nijk}
    \mathsf{N}_{ijk} = \mathsf{Q}_{ijk} \overline{a_i a_j a_k},
\end{equation}
defining the the average nonlinear transfer rate between triads of modes. The study of the organisation and structure of this tensor and how these are altered by the sparsification algorithm is the main focus of this work.

\subsection{Subspace Rotation Technique}
To identify the new set of modal structures, we utilise a subspace rotation technique introduced in the context of stabilisation of Galerkin-based reduced order models by \citet{ balajewicz2013low} (see also \citet{amsallem2012stabilization}). Geometrically, this technique consists in seeking a rotation of an $N$-dimensional POD subspace within a larger POD subspace of dimension $M$. The rotation is defined by a transformation matrix $\mathsfbi{X} \in \Re^{M \times N}$, satisfying $\mathsfbi{X}^\top \mathsfbi{X} = \mathsfbi{I}$ to ensure that the rotated spatial basis functions remain an orthonormal set. 
The rotated basis functions and the associated temporal coefficients, denoted in what follows with a tilde, are expressed as a linear combination of the original POD spatial and temporal modes as
\begin{subequations}\label{eq:new_basis}
\begin{align}
    \displaystyle \tilde{\boldsymbol \phi}_i(\mathbf{x}) =  \sum_{j=1}^{M} \mathsf{X}_{ji} {\boldsymbol \phi}_j(\mathbf{x}), \\
    \displaystyle \tilde{a}_i(t) =  \sum_{j=1}^{M} \mathsf{X}_{ji} {a}_j(t).
\end{align}
\end{subequations}
It is worth pointing out that finding a new set of modal structures directly would be a much higher dimensional problem to tackle. The number of unknowns would be proportional to the numbers of modes sought for multiplied by the number of degrees of freedom of the problem at hand. Seeking new modal structures as a linear combination of POD modes represents a significant reduction in complexity, controllable by varying the dimension $M$. Using POD modes as building blocks also has the advantage of producing a basis with good energy reconstruction properties. 

The linear and quadratic coefficients of the Galerkin system \eqref{eq:galerkin} obtained by projection on the rotated subspace are then given by the following matrix expressions
\begin{equation}
\mathsfbi{\Tilde{C}} = \mathsfbi{X}^\top \mathsfbi{C} \quad \text{and}\quad
\mathsfbi{\Tilde{L}} = \mathsfbi{X}^\top \mathsfbi{L} \mathsfbi{X},
\end{equation}
while the quadratic coefficients are cubic polynomial functions of the entries of the rotation matrix
\begin{equation}\label{eq:rotation}
    \mathsf{\Tilde{Q}}_{ijk} = \sum_{p, q, r=1}^M\mathsf{Q}_{pqr} \mathsf{X}_{pi}\mathsf{X}_{qj}\mathsf{X}_{rk},\quad i,j,k = 1, \ldots, N.
\end{equation}
In these expressions, the tensors $\mathsfbi{C} \in \Re^{M}$, $\mathsfbi{L} \in \Re^{M\times M}$ and $\mathsfbi{Q} \in \Re^{M\times M \times M}$ are the Galerkin coefficient tensors obtained from the $M$-dimensional set of original POD modes.
%
Our goal is to seek a rotation matrix for which the rotated quadratic interaction coefficient tensor $\mathsfbi{\tilde{Q}}$ has a sparse structure, i.e. where as many as possible of the quadratic interaction coefficients are identically zero. 

At this stage, it is worth noting that any rotation is necessarily accompanied by a loss of fluctuation kinetic energy reconstructed by the new basis. The energy reconstructed by a set of $P$ POD modes can be quantified by utilising the average modal energies \mbox{$\lambda_i = \overline{a_i a_i},\; i = 1, \ldots, P$}. Arranging them into a diagonal matrix ${\boldsymbol \Lambda}_P \in \Re^{P\times P}$, the trace $Tr( \boldsymbol{\Lambda}_P )$ defines an upper bound for the reconstructed energy for any $P$-dimensional set of basis functions, due to well-known optimality properties of POD. Similarly, the energy reconstructed by the rotated basis can be expressed with the average modal energies $\tilde{\lambda}_i = \overline{\tilde{a}_i \tilde{a}_i},\; i=1, \ldots N$, of the rotated temporal coefficients (\ref{eq:rotation}b) and arranging them into the diagonal matrix  $\boldsymbol{\Tilde{\Lambda}}_N = \mathsfbi{X}^\top \boldsymbol{\Lambda}_M \mathsfbi{X} \in \Re^{N\times N}$. The loss of reconstructed fluctuation kinetic energy with respect to an $N$-dimensional POD subspace is then quantified as
\begin{equation}\label{eq:objective}
    \mathcal{J}(\mathsfbi{X}) = Tr( \boldsymbol{\Lambda}_N - \mathsfbi{X}^\top \boldsymbol{\Lambda}_M \mathsfbi{X}).
\end{equation}
As observed in \citet{balajewicz2013low}, this quantity is necessarily non-negative due to the optimality of the original POD basis, i.e.
\begin{equation}
    \overline{\int_\Omega \|\mathbf{u}^\prime(t, \mathbf{x})\|^2 \mathrm{d}\Omega} = Tr(\boldsymbol{\Lambda}_\infty) > Tr(\boldsymbol{\Lambda}_N) \ge Tr(\tilde{\boldsymbol{\Lambda}}_N),
\end{equation}
where the last equality holds for $M>N$ in trivial cases only. In addition, the quantity \eqref{eq:objective} is also always identically zero when $M=N$, for any $\mathsfbi{X}$, since any linear combination of $N$ POD modes necessarily spans the same original $N$-dimensional subspace. 
%

To measure sparsity of the quadratic interaction coefficient tensor, we use the $l_1$ norm operator, denoted as $\|\cdot\|_1$, in light of the practical intractability of the zero norm in optimisation \citep{jovanovic2014sparsity}. Then, the trade-off between energy optimality and sparsity is expressed by formulating the following constrained optimisation problem

\begin{subequations}
\label{eq:optimisation_problem_P1}
\begin{alignat}{2}
&\!\min_{\mathsfbi{X}}    &\qquad& Tr(\boldsymbol{\Lambda}_N - \mathsfbi{X}^\top \boldsymbol{\Lambda}_M \mathsfbi{X}) \label{eq:optProb_a}\\
&\text{subject to} &      & ||\mathsfbi{\Tilde{Q}}||_1 \leq ||\mathsfbi{Q}||_1 / \xi,\label{eq:optProb_b}\\
&           &      & \mathsfbi{X}^\top \mathsfbi{X} = \mathsfbi{I}_{N\times N} ,\label{eq:optProb_c}
\end{alignat}
\end{subequations}
referred to as problem $\text{P1}$ in what follows. The role of the objective function \eqref{eq:optProb_a} is to favour transformation matrices that minimise the loss of energy optimality, producing a set of basis function with good energy representation ability, as in \citet{balajewicz2016minimal}. On the other hand, the constraint \eqref{eq:optProb_b} encourages sparse solutions, because some entries of $\mathsfbi{\tilde{Q}}$ are shrunk to zero during the solution of (\ref{eq:optimisation_problem_P1}) by the non-differentiability of the $l_1$ norm \citep{tibshirani2013lasso, stat_learning}. Then, the weight $\xi$ is an arbitrary penalisation parameter that controls the relative $l_1$ norm of the rotated quadratic coefficient tensor and the sparsity of the resulting Galerkin model.
It is worth noting that, unlike in the \emph{a-posteriori} LASSO-based sparsification methods \citep{brunton2016discovering, rubini2020l1}, not all quadratic coefficients can be simultaneously set to zero by an arbitrary rotation. In fact, the tensor $\mathsfbi{\tilde{Q}}$ depends nonlinearly on the rotation $\mathsfbi{X}$, while in LASSO-based methods the tensor coefficients are directly the optimisation variables of the problem 
This indicates that problem \eqref{eq:optimisation_problem_P1} might not have a feasible solution if the penalisation weight is too large.

An important characteristic of optimisation problem \eqref{eq:optimisation_problem_P1} is that while the objective (\ref{eq:optProb_a}) is convex, the sparsity-promoting constraint (\ref{eq:optProb_b}) is not, as it involves cubic polynomials in the optimisation variables, the entries of the transformation matrix $\mathsfbi{X}$. Consequently, the solution might no be unique and several local minima, corresponding to different sets of basis functions, may be obtained by starting the optimisation from different initial guesses. However, as demonstrated in appendix \ref{app:uniqueness}, starting the optimisation from small random perturbations of the original POD basis (justified by the need to retain good energy reconstruction properties) consistently produced the same optimal solutions, which will be presented in section \ref{sec:results}.

Assuming a feasible solution of problem \eqref{eq:optimisation_problem_P1} can be found, the Galerkin model constructed from projection onto the optimal rotated basis does not necessarily possess better long-term temporal stability characteristics than the original POD model. 
In fact, it is well-known that POD-Galerkin models exhibit long-term instability because of the deficit of energy dissipation attributed to the truncation of small dissipative scales \citep{noack2008finite, schlegel_noack_2015}. In the present case, the transformation $\mathsfbi{X}$ obtained from solution of \eqref{eq:optimisation_problem_P1} does not necessarily result in an improved description of dissipative processes. Classically, this issue is cured by introducing, \emph{a posteriori}, an eddy-viscosity-type term in the Galerkin model \citep{galletti2004low, noack2005need,osth2014need}. However, an \emph{a posteriori} correction would not remain in the spirit of the present work. We thus favour the subspace-rotation-based stabilisation approach proposed by \citet{balajewicz2013low, balajewicz2016minimal}, which can be introduced naturally in the present formulation. In practice, we augment problem \eqref{eq:optimisation_problem_P1} with the additional implicit constraint
\begin{equation}
    Tr(\tilde{\mathsfbi{L}}) = Tr(\mathsfbi{X}^\top \mathsfbi{L} \mathsfbi{X}) = -\eta
\end{equation}
where the auxiliary variable $\eta \in \Re^+$ is chosen such that
\begin{equation}\label{eq:chi}
    \chi(\eta) = \frac{\overline{E(t)}-\overline{E_\text{DNS}(t)}}{\overline{E_\text{DNS}(t)}} = 0,
\end{equation}
i.e.~that the relative difference of the average fluctuation kinetic energies from direct numerical simulation (DNS) and from numerical simulation of the new model vanishes. The variable $\eta$ controls dissipation mechanisms in the Galerkin model by altering the spectrum of $\mathsfbi{\tilde{L}}$ and ensures long-term stability. As observed by \citet{balajewicz2013low}, $\eta$ is not known a priori, but can be found in an inner optimisation loop to ensure that the excess of turbulent kinetic energy defined by $\chi(\eta)$ is zero. 
With this additional constraint, problem \eqref{eq:optimisation_problem_P1} becomes
\begin{subequations}
\label{eq:optimisation_problem_P2}
\begin{alignat}{2}
&\!\min_{\mathsfbi{X}}    &\qquad& Tr(\boldsymbol{\Lambda}_N - \mathsfbi{X}^\top \boldsymbol{\Lambda}_M \mathsfbi{X}) \label{eq:optProb_2a}\\
&\text{subject to} &      & ||\mathsfbi{\Tilde{Q}}||_1 \leq ||\mathsfbi{Q}||_1 / \xi,\label{eq:optProb_2b}\\
&           &      &Tr(\mathsfbi{X}^\top \mathsfbi{L} \mathsfbi{X}) = - \eta \quad \text{with} \quad \chi(\eta) = 0,\label{eq:optProb_2c}\\
&           &      & \mathsfbi{X}^\top \mathsfbi{X} = \mathsfbi{I}_{N\times N}.\label{eq:optProb_2d}
\end{alignat}
\end{subequations}
In this formulation, denoted as P2 henceforth, there is still only one free parameter, $\xi$. The additional constraint guarantees long-term stability but it can only be satisfied (i.e. the problem is feasible) if $M > N$. In fact, any rotation $\mathsfbi{X} \in \Re^{N\times N}$ cannot alter the subspace spanned by the original $N$ POD modes and the associated description of dissipation mechanisms captured  by the model. In what follows, we consider models with ratio $M/N = 2$ and 3. 

In practice, a small random perturbation of the original $N$-dimensional POD basis was used as initial guess for the optimisation and successive optimisation problems for different penalisations $\xi$ were started from the solution of the previous problem. Problem \eqref{eq:optimisation_problem_P2} was solved with the open-source package for nonlinear and non-convex optimisation \texttt{NLopt} \citep{johnson2014nlopt}. We utilised a solver implementing the Method of Moving Asymptotes (MMA) algorithm \citep{Svanberg_aclass} which requires the gradient of the objective function and of the constraints. The key element to make the procedure viable is to evaluate the sparsity-promoting constraint and its gradient as efficiently as possible. A naive implementation requires $\mathcal{O}(M^3 N^3)$ operations for the evaluation of the sparsity promoting constraint and $\mathcal{O}(M^4 N^4)$ for the evaluation of its gradient with respect to the rotation $\mathsfbi{X}$, and costs become quickly intractable. A significantly more efficient algorithm to compute these two quantities with costs scaling as \mbox{$\mathcal{O}(M N^3 + M^2 N^2 + M^3 N)$} for the evaluation of the sparsity-promoting constraint and its gradient can be derived and is key to make the procedure viable (see discussion of the method in appendix \ref{app:comp_cost}).


One further technical remark is that the sparsity-promoting constraint is a non-smooth function of the transformation matrix $\mathsfbi{X}$, posing difficulties for the utilisation of gradient-based optimisation algorithms. One approach is to implement a subgradient descent method, often used for the solution of the LASSO \citep{stat_learning}. However, this algorithm can only be implemented if an analytical solution of the optimisation problem is known, which is not the case here. In this work, we used a manual soft-thresholding approach where entries of the rotated tensor $\mathsfbi{\tilde{Q}}$ smaller that the numerical tolerance specified to the gradient-based optimiser (typically $tol = 10^{-5}$) are set to zero at the end of the optimisation. 

The Galerkin models obtained from solution of \eqref{eq:optimisation_problem_P2} are then characterised by examining the density of the rotated triadic interaction tensor, defined as 
\begin{equation}\label{eq:density}
    \rho = \frac{||\mathsfbi{\Tilde{Q}}||_0}{||\mathsfbi{Q}||_0}, 
\end{equation}
where the $l_0$ norm $||\cdot||_0$ counts the nonzero elements of a tensor. The density can also be expressed as the average 
\begin{equation}
    \label{eq:average_modal_density}
    \rho = \frac{1}{N} \sum_{i=1}^N \rho_i 
\end{equation}
with the modal densities $\rho_i = ||\mathsfbi{\Tilde{Q}}_i||_0 /||\mathsfbi{Q}_i||_0$, $i =1, \ldots, N$ being the relative number of non-zero coefficients in the slices $\mathsfbi{Q}_i$ of the quadratic coefficient tensor associated to each modal index. To express the energy captured by the $N$-dimensional rotated basis, we also introduce the global energy reconstruction factor
\begin{equation}
    e_N = Tr(\tilde{\boldsymbol{\Lambda}}_N) / Tr(\boldsymbol{\Lambda}_\infty),
    \label{eq:energy_reconstructed}
\end{equation}
a quantity always strictly lower than one.  The density and the reconstruction factor depend on the dimensions $M$ and $N$ and on the penalisation weight $\xi$. To characterise the effects of these parameters, we visualise the rotated systems on the $\rho-e_N$ plane, similar to the approach used in \citet{rubini2020l1}.

\section{Demonstration: two-dimensional lid-driven unsteady cavity flow}\label{sec:results}
We now apply this methodology to two-dimensional unsteady flow in a lid-driven square cavity. This is an established test case for the development and validation of model order reduction techniques \citep{cazemier1998proper, terragni, balajewicz2013low, arbabi2017study, fick2017reduced}, and we thus consider it here as an exemplar to demonstrate the ideas discussed in the introduction. This is the same test case we utilised in our previous work \citep{rubini2020l1} to demonstrate the properties of $l_1$-based sparsification of Galerkin models.

\subsection{Problem definition and Proper Orthogonal Decomposition}
The Reynolds number is defined as $Re = LU/\nu$ where $L$ and $U$ are the cavity dimension and the (uniform) lid velocity, respectively, and $\nu$ is the kinematic viscosity. All physical variables introduced later on are scaled with $L, U$ and combinations thereof. We consider the flow regime establishing at $Re = 2 \times 10^4$, where the motion is chaotic \citep{auteri2002numerical, PENG2003337}. The domain is defined by the nondimensional Cartesian coordinates $\mathbf{x} = (x, y)$, and the velocity vector $\mathbf{u}(t, \mathbf{x})$ is defined by the components $u(t, \mathbf{x})$ and $v(t, \mathbf{x})$. For visualisation purposes, we introduce the out-of-plane vorticity component $\omega = \partial v/\partial x - \partial u/\partial y$.
\begin{figure}
  \centerline{\includegraphics[width=1\textwidth]{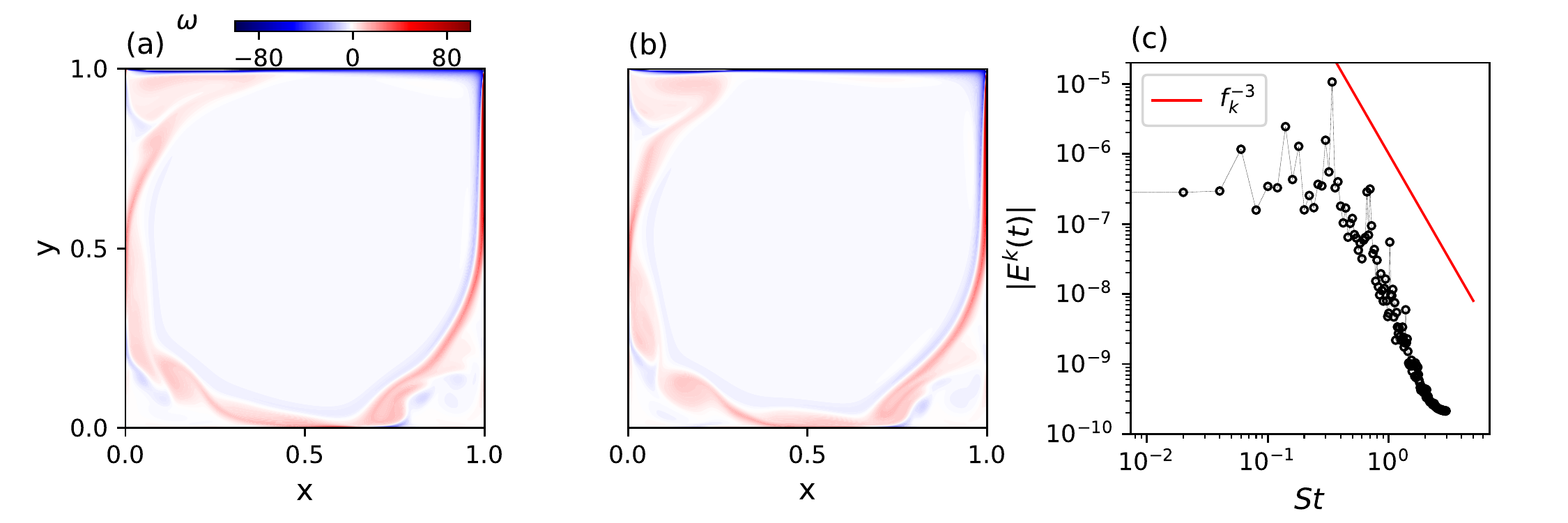}}
  \caption{Instantaneous vorticity fields for two different flow snapshots in panels (a) and (b). Amplitude of the Fourier transform of the fluctuating kinetic energy signal (equation \ref{eq:energy_cont}),  in panel (c).}
  \label{fig:vorticity_snapshots}
\end{figure}
Numerical simulations were performed in \texttt{OpenFOAM} with the unsteady incompressible flow solver \texttt{icofoam}. The convective and viscous terms are spatially discretised with a second-order finite-volume technique and the temporal term with a semi-implicit Crank-Nicholson scheme. 
A grid independence study was performed, examining average and unsteady flow quantities on increasingly finer meshes. The final mesh is composed of $300 \times 300$ cells, with refinement at the four cavity boundaries. This mesh is sufficiently fine to resolve the unsteady high-shear regions bounding the main vortex, the high vorticity filaments characteristic of two-dimensional turbulence as well as the spatial structure of the lowest energy POD modes utilised in this study. Similar grid resolutions have been used by \cite{cazemier1998proper} at similar Reynolds numbers. 


Two snapshots of the vorticity field obtained from these simulations are shown in figure \ref{fig:vorticity_snapshots}-(a,b). Most of the dynamically interesting features in this regime originate at the bottom-right corner of the cavity. Specifically, the secondary vortex in the recirculation zone is shed erratically, producing wave-like disturbances advected along the shear layer bounding the primary vortex. These structures produce a strong quasi-periodic oscillation of the turbulent kinetic energy, as shown in panel (c), where the amplitude of the Fourier transform of the fluctuation kinetic energy signal (equation (\ref{eq:energy_cont})) is shown, as a function of the Strouhal number $St = f L / U$, with $f$ the dimensional frequency. Due to the chaotic nature of the flow, the energy spectrum has a strong broadband component with energy uniformly distributed across all the frequencies/spatial structures of the flow. This implies that unlike for periodic flows \citep{noack2011reduced, symon2020energy}, the mean triadic interaction tensor $\mathsfbi{N}$ is dense with energy transfers highly scattered in modal space. However, the wave-like motion characterising the shear layer dynamics produces a strong energy peak at a characteristic non-dimensional Strouhal number $St = 0.7$ and its harmonics.

From these simulations, we extract $N_T = 1000$ snapshots after initial transients have decayed, using a nondimensional sampling period $\Delta t = 0.1$. These settings are sufficient to adequately time-resolve the fastest scales of motion as well as to include many shedding events originating at the bottom right corner. The snapshot Proper Orthogonal Decomposition of this dataset is then performed \citep{sirovich1987turbulence}. The normalised cumulative reconstructed energy  \eqref{eq:energy_reconstructed} is reported in table \ref{tab:energy_pod} as a function of the mode number $N$.
\begin{table}
  \begin{center}
\def~{\hphantom{0}}           
  \begin{tabular}{lllllllllllll}
       $N$  & 1 & 5 & 10 & 20 & 30 & 40 & 60 & 80 & 90  & 120  \\[3pt]
       \hline
       $e_N$   & 0.26 & 0.74 & 0.9 & 0.96 & 0.97 & 0.98 & 0.985& 0.99 & 0.995 & 0.998     
  \end{tabular}
  \caption{Normalised cumulative energy reconstruction of the original POD basis functions.}
  \label{tab:energy_pod}
  \end{center}
\end{table}

\subsection{Analysis of energy interaction of the original POD-Galerkin models}
Before examining the properties of the sparsified Galerkin models, we first present some aspects of energy transfers of Galerkin models constructed using the original POD modes, using a reference model size $N=30$.
\begin{figure}
  \centerline{\includegraphics[width=1.05\textwidth]{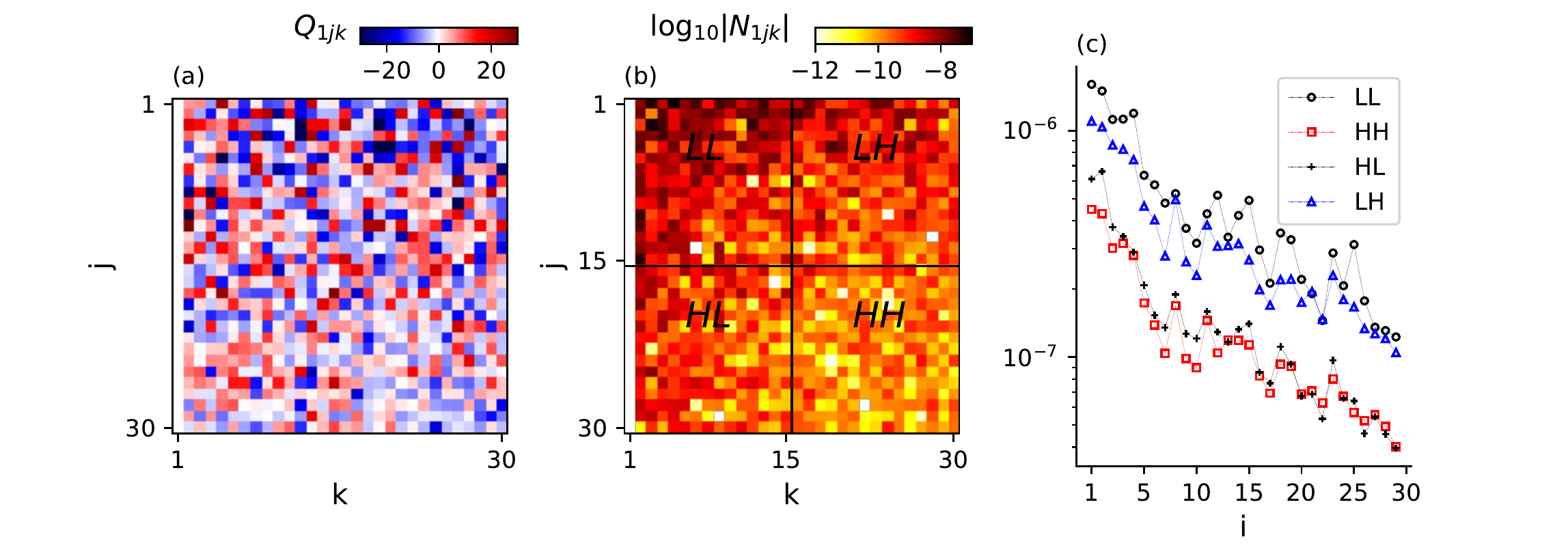}}
  \caption{Organisation of the quadratic coefficients $\mathsfbi{Q}$ and of the average nonlinear transfer rate $\mathsfbi{N}$ for mode $i=1$, in panels (a) and (b), respectively. 
  Collective energy transfer rates associated to the four regions defined in panel (b), as a function of the modal index $i$. }
  \label{fig:Energy_Analsis}
\end{figure}
Figure \ref{fig:Energy_Analsis}-(a) shows the organisation of the quadratic coefficients $\mathsfbi{Q}$. The slice for mode $i=1$ is used as an illustrative example; other modes have similar characteristics. The coefficient tensor is dense as most coefficients are nonzero. No clear structure in the distribution of the coefficients can be observed, except for a slight asymmetry where coefficients for $k>j$ are often larger in magnitude, for mode $i=1$ but also for most of the other modes. This asymmetry is a consequence of the definition of the quadratic coefficient tensor \eqref{eq:definition_qijk}, characterising convective transport mechanisms, and the range of spatial length scales described by the POD modes, as already observed in \citet{rubini2020l1}. 
One slice of the average energy transfer rate tensor $\mathsfbi{N}$ is reported in panel (b) using the base ten logarithm of the magnitude. All entries of $\mathsfbi{N}$ are generally nonzero, although the inter-modal transfers in the Galerkin model are highly organised and the intensity of interactions varies across several orders of magnitude. This is the combined result of the structure of the quadratic coefficients tensor and of the complex spectral structure of the temporal coefficients.  

To further characterise energy paths, we split the interactions in four regions, denoted as LL, LH, HL and HH, with L and H denoting low and high index modes, respectively. These four regions represent a coarse-grained grouping of energy transfers between scales resolved by the model, assuming that low index modes map to the largest scales of motion in the cavity and high index modes describe small-scale, low-energy features. Here, we arbitrarily select the cut-off at half of the spectrum ($N=15$), but other choices are possible and do not change the following results qualitatively. 
We then compute the sum of the magnitude of the average transfer rates contained in these four regions for each mode, to characterise in a coarse-grained fashion energy transfers between triads of high-energy modes and smaller, dissipative scales. The result of this analysis is shown in figure \ref{fig:Energy_Analsis}-(d), as a the modal index $i$.  First, energy transfers in the regions LL, HL and LH are generally more intense than those in region HH. This follows from the observation that large-scale/large-scale and large-scale/small-scale interactions are more relevant with respect the small-scale/small-scale interactions, across the entire hierarchy. In addition, we observe that the interactions LH are always more intense than the HL interactions. This is consequence of the asymmetry previously observed in the tensor $\mathsfbi{Q}$ and not of the temporal coefficients, due to definition \eqref{eq:Nijk}, where the indices $j$ and $k$ commute. This lack of symmmetry is in agreement with the picture of energy transfers between scales in homogeneous isotropic two-dimensional turbulence \citep{ohkitani1990nonlocality,laval1999nonlocality}, where the large scales interact with the small ones in a non-local fashion.

\subsection{Model sparsification}

\begin{figure}
  \centerline{\includegraphics[width=0.92\textwidth]{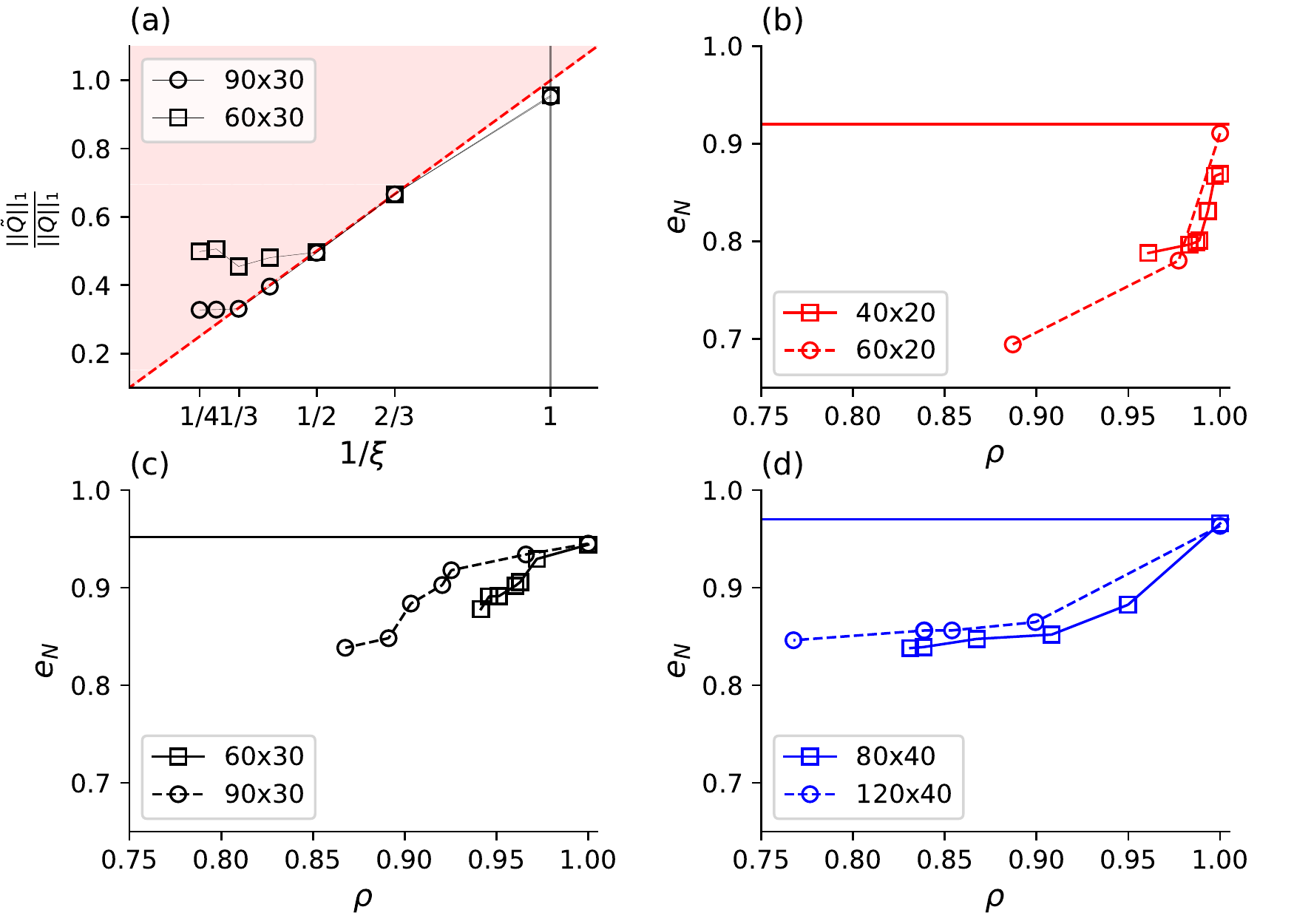}}
  \caption{Panel (a): visualisation of the sparsity-promoting constraint \eqref{eq:optProb_2b} on the plane \mbox{($1/\xi$ -- $||\mathsfbi{\tilde{Q}}||_1/||\mathsfbi{Q}||_1$)}. The red region denotes the infeasibity set. Panels (b) to (d): $\rho-e_N$ curves for families of models with dimension $N=30, 20, 40$ and ratios $M/N = 2, 3$ (open squares and open circles, respectively). The labels in the legend are defined with key $M \times N$.}
  \label{fig:EnegyDensity_feasibility}
\end{figure}


We now consider models with dimension $N=30$, for ratios $M/N = 2$ and $3$ and examine in more detail the effect of the sparsity-promoting constraint \eqref{eq:optProb_2b}.
For each ratio, a family of models with different density and reconstructed kinetic energy is generated by increasing the penalisation parameter $\xi$. Optimal solutions are displayed on the plane $1/\xi$ -- $||\mathsfbi{\tilde{Q}}||_1/||\mathsfbi{Q}||_1$ in figure \ref{fig:EnegyDensity_feasibility}-(a). The red dashed line separates solutions that satisfy the sparsity-promoting constraint \eqref{eq:optProb_2b} (white area feasibility region) from solutions that do not (red area). Note that the stability constraint is satisfied for all points reported in this figure, as we have noted that the optimiser is still able to satisfy \eqref{eq:optProb_2c} when it first fails to satisfy the sparsity-promoting constraint \eqref{eq:optProb_2b}.

For small penalisation weights, the optimisation problem has feasible solutions that fall on the boundary of the feasibility region. This suggests that rotations of the original POD basis that minimise the energy loss are found on the boundary, i.e. that there is a competition between sparsification and energy representation. The key feature of figure \ref{fig:EnegyDensity_feasibility}-(a) is that there exists a threshold value $\xi_t$ above which the optimisation problem terminates unsuccessfully in the unfeasible region, i.e. no rotation exists that can reduce the $l_1$ norm of the rotated quadratic coefficient tensor below $\|\mathsfbi{Q}\|_1/\xi_t$. This is manifested in panel (a) by a sudden turn of the solution traces from the feasible region boundary upwards into the red region. This occurs because, for a given dimension $M$, there is only a limited number of coefficients in the tensor $\tilde{\mathsfbi{Q}}$ that can be shrunk to zero by any rotation of the basis functions. The threshold value increases with the ratio $M/N$, i.e.~the $l_1$ norm of the rotated coefficient tensor $\tilde{\mathsfbi{Q}}$ can be decreased further when larger dimensions $M$ are used. This is arguably a consequence of the fact that higher ratios $M/N$ corresponds to more degrees of freedom available to the optimiser to ensure that the sparsity constraint is satisfied. 
It is worth noting that, in LASSO-based sparsification methods \citep{rubini2020l1}, the optimiser operates directly on the model coefficients and feasible solutions can always be found, with all coefficients shrunk to zero in the limiting case. However, highly sparsified models were observed to have little physical significance and poor temporal behaviour. 
Here, all feasible systems with varying sparsity/energy reconstruction properties are temporally stable and provide physically consistent predictions. 



In panels (c, b d) of figure \ref{fig:EnegyDensity_feasibility} the trade-off between sparsity and the energy reconstruction properties of the rotated basis is presented on the $\rho-e_N$ plane for models constructed with $N=20, 30, 40$, respectively. The horizontal line in each panel corresponds to the fraction of reconstructed energy of the original dense POD-Galerkin model. In these panels, squares are used to denote data for $M/N = 2$, while circles denote data for $M/N=3$. It can be first observed that systems for $\rho = 1$ ($\xi=1$) do not reconstruct the entire fluctuation kinetic energy captured by the original POD basis. This is due to the stability constraint \eqref{eq:optProb_2c}, producing a small rotation of the optimal POD basis so that dissipative mechanisms in the Galerkin models to ensure long-term stability are better resolved. Second, when the penalisation $\xi$ is increased, sparser models are obtained, with higher ratios $M/N$ enabling further reduction in density. This, however, comes at the cost of decreasing the energy optimality of the rotated basis. More importantly, the larger the model size, the more the model can be sparsified without significantly affecting the ability of the new basis to reconstruct the fluctuation kinetic energy. This suggests that the sparsification technique becomes more effective as the model complexity and the range of scales resolved by the basis, increases. This appears to be a general trend, since a similar behaviour was observed in \citet{rubini2020l1} using the LASSO-based \emph{a-posteriori} sparsification. 



\subsection{Analysis of the rotated modal structures}
\begin{figure}
  \centerline{\includegraphics[width=1\textwidth]{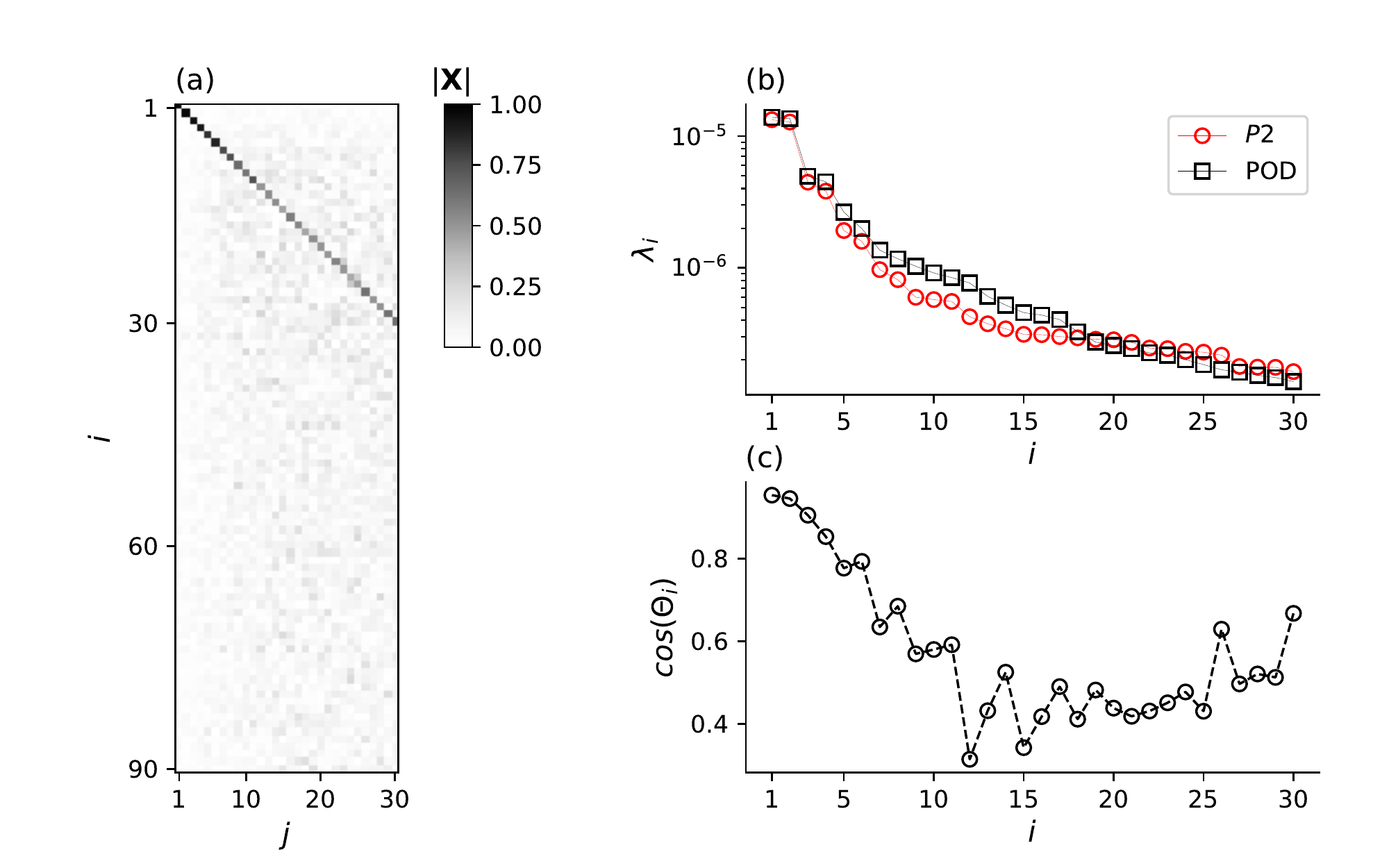}}
  \caption{Optimisation results for the case $M/N=3$, $N=30$, $\xi=3$ (with density $\rho = 0.87$). Panel (a): magnitude of the entries of the transformation matrix $\mathsfbi{X}$. Panel (b): distribution of the average modal energies of the original (square) and rotated (circles) basis functions. Panel (c): cosine of the angle between the modes of the original and rotated basis.}
  \label{fig:New_modal_basis}
\end{figure}

We now move to the analysis of the rotated spatial and temporal basis functions.
A model with $N=30$, $M/N = 3$ and density $\rho=0.87$, obtained for $\xi=3$ just before the solution falls into the infeasible region in figure \ref{fig:EnegyDensity_feasibility}-(a) is analysed.  Panel (a) of figure \ref{fig:New_modal_basis} shows the magnitude of the entries of the rotation matrix $\mathsfbi{X}$, found from the solution of \eqref{eq:optimisation_problem_P2}. Panel (b) of the same figure compares the modal energies of the original POD temporal coefficients with those of the rotated modes. Panel (c) shows the cosine of the angle between each pair of original and rotated spatial modes, $\cos(\theta_i) = (\boldsymbol{\phi}_i,  \tilde{\boldsymbol{\phi}}_i)$, which is clearly also the diagonal of $\mathsfbi{X}$, because of the orthogonality constraint \eqref{eq:optProb_2d}.
The transformation matrix $\mathsfbi{X}$ has large diagonal entries, but significant off-diagonal terms can be observed for $i, j \gtrsim 5$. This indicates that the rotated basis functions bear a strong resemblance to the original POD modes, but that the optimisation has introduced into the new basis small-scale, low-energy features to both stabilise and sparsify the rotated Galerkin model. It is observed that the high-energy temporal and spatial modes are not significantly affected by the rotation and do not differ significantly from the original POD modes. For instance, the first pair of modal energies, corresponding to the dominant fluid oscillation in the cavity, is virtually unchanged. Conversely, high-index, low-energy modes are more significantly rotated away from the corresponding original POD mode and more significant relative differences of the modal energies are observed. We argue that this behaviour derives from the formulation of problem \eqref{eq:optimisation_problem_P2}, constructed with the aim of generating a basis that minimises the energy loss with respect to the energetically optimal POD. As a result, the optimisation leaves mostly unchanged the most energetic modes that contribute more pronouncedly to the overall energy and rotates by a larger extent the less energetic modes to gain in sparsity and to achieve stability.
\begin{figure}
  \centerline{\includegraphics[width=1.1\textwidth]{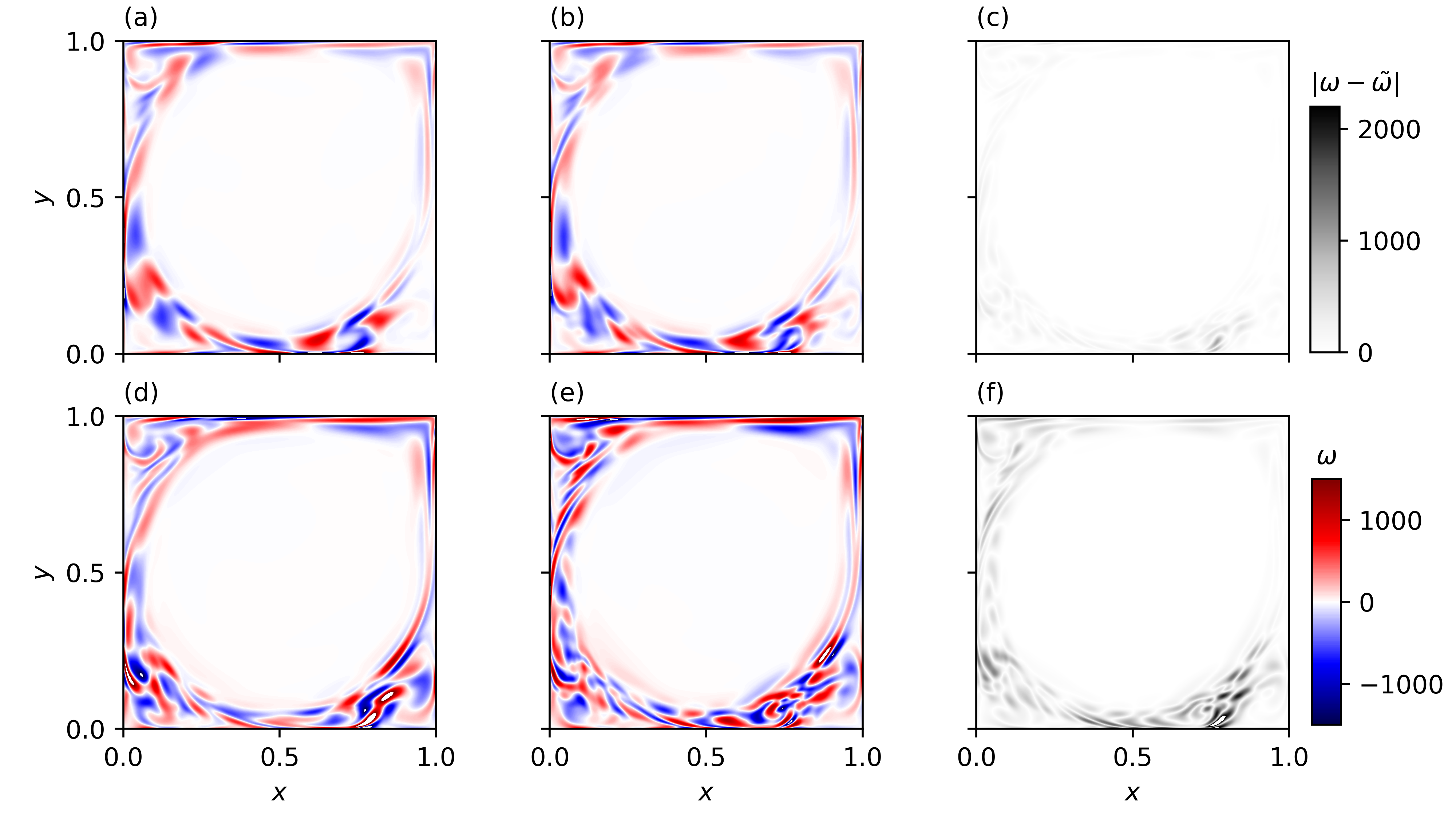}}
  \caption{Vorticity fields of the first original POD mode, panel (a), the first rotated mode (denoted as $\tilde{\omega}$), panel (b), and the absolute value of their difference, panel (c). Panels (d), (e) and (f) show the same quantities for mode $i = 19$.}
  \label{fig:Spatial_modes}
\end{figure}

Figure \ref{fig:Spatial_modes} shows the out-of-plane vorticity component $\omega$ for the original POD spatial mode, panels (a,d), and the rotated mode, panels (b,e), for indices $i=1$ and $19$, respectively. The third column, panels (c,f), shows the absolute value of the difference between the original and rotated modes. As expected, the spatial structure of the first mode is not considerably changed by the rotation. On the other hand, mode $i=19$ is more pronouncedly affected by the rotation, with small-scale vorticity features appearing all along the shear layer. We argue that the introduction of small scale features is a combined effect of the stability constraint \eqref{eq:optProb_2c}, which enhances dissipation in the system \citep{balajewicz2016minimal}, and of the sparsity-promoting constraint \eqref{eq:optProb_2b} since spatially fluctuating modes, with stronger gradients, are likely more effective in shrinking to zero the spatial averages involved in the projection coefficients \eqref{eq:definition_qijk}.
\begin{figure}
  \centerline{\includegraphics[width=1\textwidth]{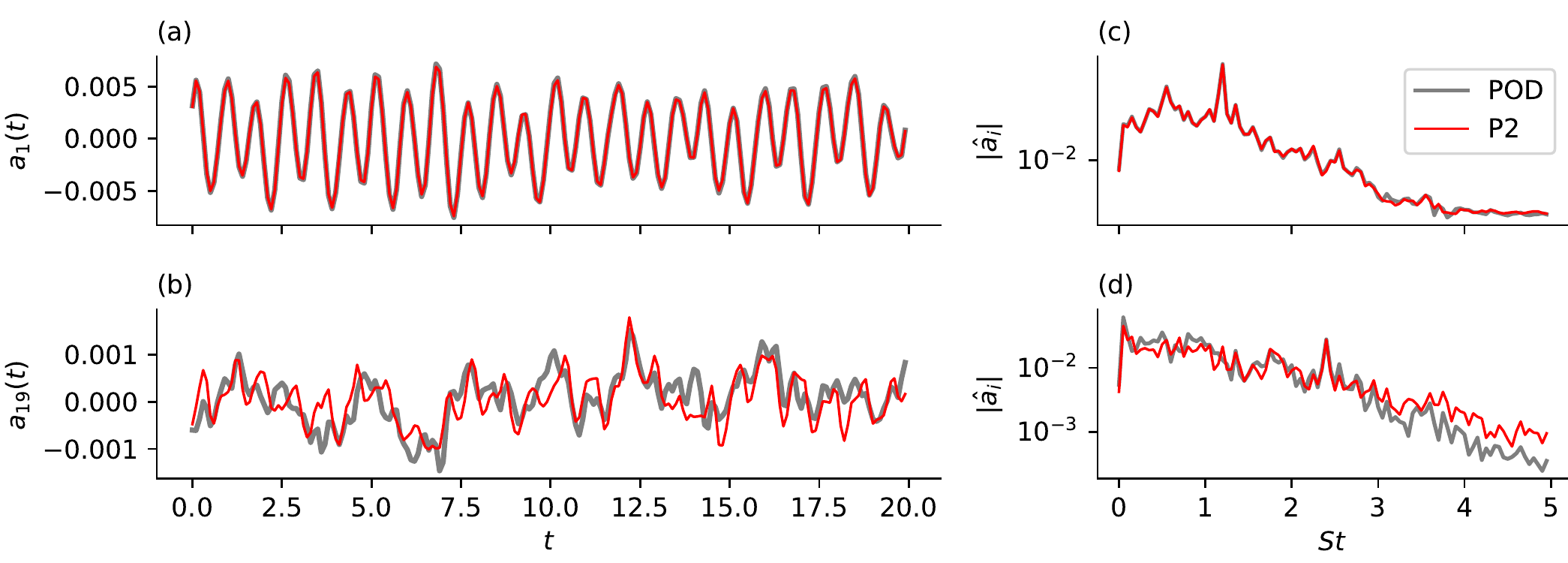}}
  \caption{Temporal evolution and amplitude spectrum of $a_1(t)$ in panels (a) and (c), respectively, and for $a_{19}(t)$ in panels (b) and (d). Data is reported for the original POD temporal modes and for the rotated modes of the sparse system obtained from formulation P2.}
  \label{fig:Temporal_modes}
\end{figure}

The temporal modes are affected in a similar way. This is illustrated in figure \ref{fig:Temporal_modes} showing the temporal evolution of modes $a_1(t)$ and $a_{19}(t)$ over twenty time units, panels (a) and (b), and their amplitude spectra in panels (c) and (d), respectively. Since the first column of $\mathsfbi{X}$ is close to zero except for $\mathsf{X}_{11}$, mode $a_1(t)$ and its spectral content is not appreciably affected by the rotation, except for a small general decrease of the amplitude due to decrease in energy content (see figure \ref{fig:New_modal_basis}-(b)). Conversely, the spectral content of mode $a_{19}(t)$ is remodulated by the rotation by introducing higher energy at high frequency components, consistent with the introduction of small-scale features into the corresponding spatial mode.

\subsection{Interactions identified in the sparse model}
The structure of the spatial and temporal modes is only weakly modified by the optimisation, but this is sufficient to introduce sparsity in the rotated quadratic coefficient tensor $\tilde{\mathsfbi{Q}}$ and in the rotated average triadic interaction tensor $\tilde{\mathsfbi{N}}$ when the domain integrals \eqref{eq:definition_qijk} and the temporal averages \eqref{eq:Nijk} are computed.

To visualise how sparsity in these tensors varies when the penalisation weight $\xi$ is increased, we introduce the tensor $\boldsymbol{\xi}$ with entries $\xi_{ijk}$ defined as the value of $\xi$ in \eqref{eq:optimisation_problem_P2} at which the corresponding coefficient $\tilde{\mathsfbi{Q}}_{ijk}$ is first shrunk to zero.
\begin{figure}
  \centerline{\includegraphics[width=1\textwidth]{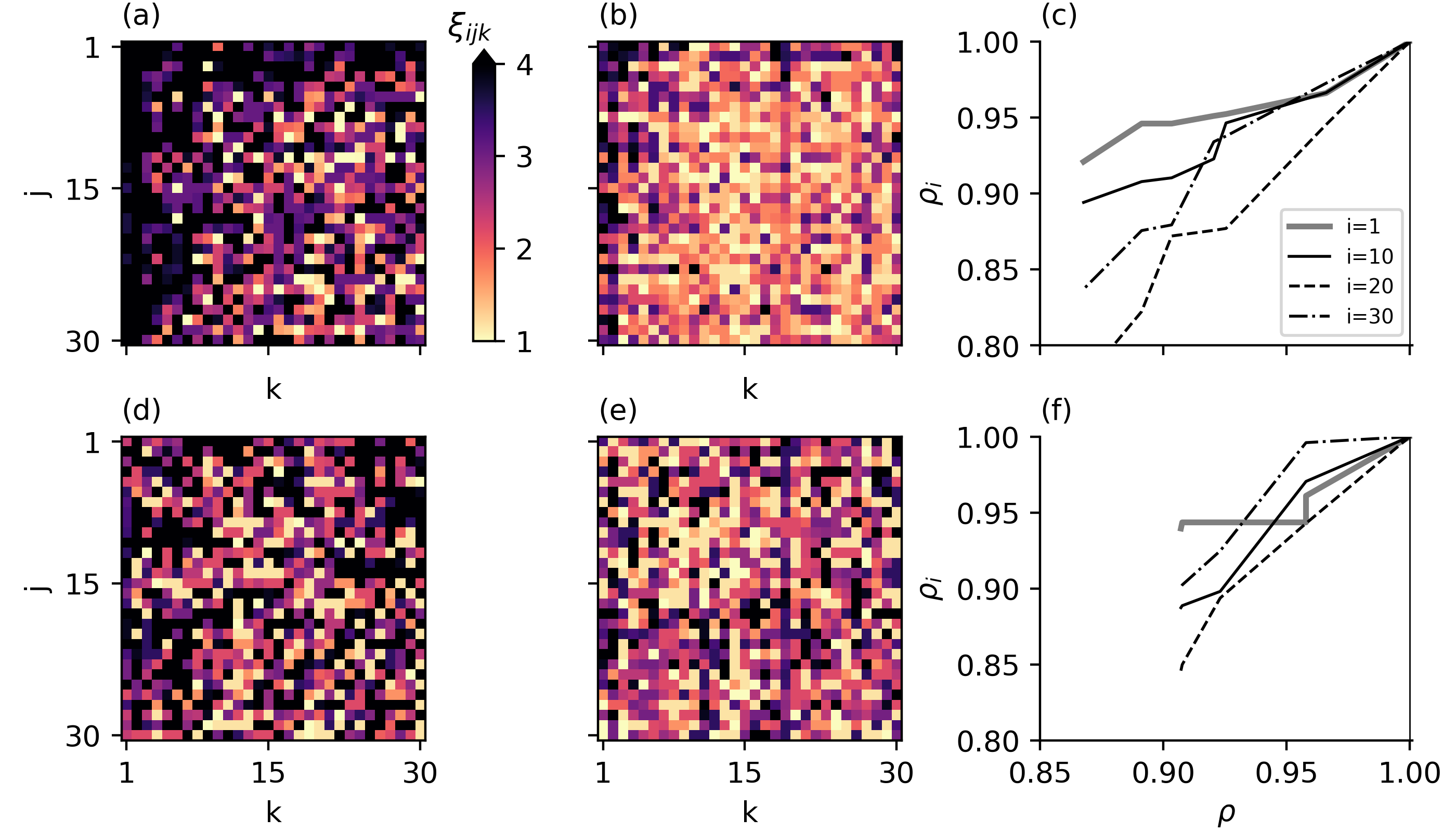}}
  \caption{Entries of the tensor $\boldsymbol{\xi}$ for $i=1$, $30$ obtained by solving problem P2, in panels (a) and (b), respectively. In panel (c): modal densities $\rho_i$ as a function of $\rho$. In panels (d), (e) and (f), the same quantities are shown from solution of the optimisation problem P1.}
  \label{fig:Sparsity_pattern}
\end{figure}
Figure \ref{fig:Sparsity_pattern} shows two slices of $\boldsymbol{\xi}$ for $i=1$ and $i=30$, for a model with $N=30$ and $M/N = 3$. Results in panels (a), (b) and (c) are obtained with the complete formulation P2, including both the stability and sparsity promoting constraint, while those in panels (d), (e) and (f) are obtained with formulation P1, which only includes the sparsity promoting constraint. Panels (c) and (f) show the modal density $\rho_i$ as a function of the global density $\rho$ for four modes across the hierarchy. By using formulation P2, which includes the stability constraint (\ref{eq:optProb_2c}), we observe that small-scale/small-scale interactions disappear first, for moderate penalisations, especially for the high-index modes, but generally across the entire hierarchy of modes. Increasing the penalisation, coefficients corresponding to interactions that are local in modal space are progressively pruned, leaving only coefficients capturing non-local interactions with the low-index modes for large penalisations. The key remark is that the structure of the sparsified quadratic coefficient tensor $\tilde{\mathsfbi{Q}}$, solution of the proposed optimisation approach, follows the same pattern displayed by triadic energy interactions shown in figure \ref{fig:Energy_Analsis}-(b), i.e. coefficients corresponding to energetically weak interactions are pruned first and only relevant interactions are preserved. Interestingly, figure \ref{fig:Sparsity_pattern}-(c) indicates that high-index modes can be sparsified more efficiently. This might be related to the fact that high-index modes are also rotated more aggressively during the optimisation, to ensure that the sparsity promoting constraint \eqref{eq:optProb_2b} is satisfied and with minor effect of the overall energy reconstruction ability.

Nonetheless, considering now the solution obtained from formulation P1 in the bottom panels of figure \ref{fig:Sparsity_pattern}, it is clear that many quadratic coefficients are indeed shrunk to zero during the optimisation and similar global densities are obtained. However, the sparsity pattern does not have a clear relation with the original structure of energy interactions. By contrast, coefficients corresponding to important energy interactions have been shrunk to zero since the optimisation problem P1 is entirely driven by the sparsity-promoting constraint.
While in both formulations the same penalisation on the $l_1$ norm of the rotated tensor $\mathsfbi{\tilde{Q}}$ is used, formulation P1 lacks any information regarding the dynamics and temporal evolution of the Galerkin model and the structure of inter-modal energy transfers. This results in an unphysical equal contraction of all quadratic coefficients, across all modes and independently of the strength of the energy interactions they represent. 
The effect of this behaviour can also be noticed in the $\rho_i-\rho$ curves, panel (f), where different individual modes are sparsified by a similar amount. Nevertheless, no significant differences are observed between models obtained from the two formulations on the $\rho-e_N$ plane.

%

\subsection{Temporal integration and energy analysis of the sparsified system}
\begin{figure}\label{fig:PDF_TKE}
  \centerline{\includegraphics[width=1\textwidth]{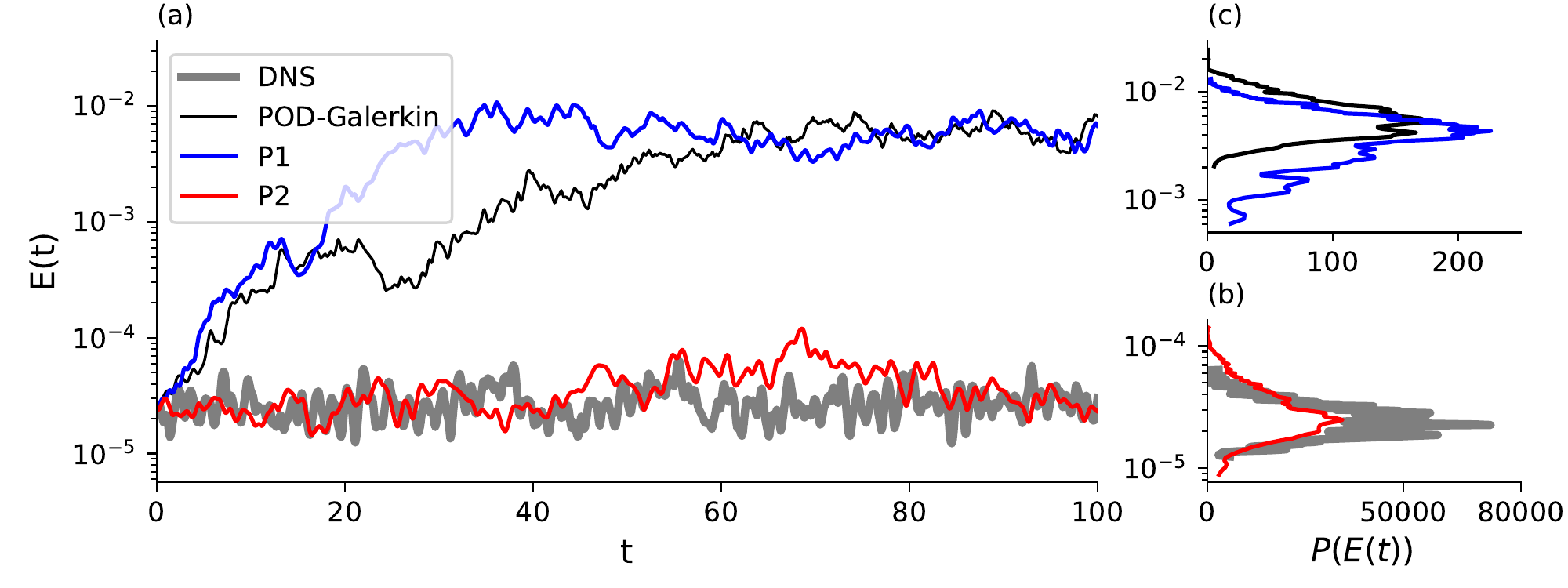}}
  \caption{Temporal evolution of the integral fluctuation kinetic energy $E(t)$. The temporal performance of the two sparse models obtained by solving problem \eqref{eq:optimisation_problem_P1} are compared against the DNS and the dense model obtained with Galerkin projection. Panels (b,c) show the probability distribution of the energy $E(t)$, obtained for long-time integration of the Galerkin models.}
\end{figure}

In this section, we analyse the temporal behaviour of the sparse reduced order models obtained by solving problems P1 and P2 and consider their rotated triadic interactions tensors $\tilde{\mathsfbi{N}}$ and the average energy budget of equation \eqref{eq:energy_avg}. The same configuration studied in previous sections, with $M/N = 3$, $N=30$ and $\xi = 3$, is considered. Models are integrated forward in time with an implicit time stepping scheme for $T = 500$ time units, from an initial condition obtained from one of the snapshots. Figure \ref{fig:PDF_TKE}-(a) shows the first one hundred time units of the temporal evolution of the turbulent kinetic energy \eqref{eq:energy_cont} for these two models, compared with the evolution from DNS and from the original POD-Galerkin model. Figure \ref{fig:PDF_TKE}-(b,c) show the probability density function of the same quantity, computed over a longer time span.
%
As expected, the deficit of dissipation in original POD-Galerkin model produces fluctuation kinetic energy levels about two orders of magnitude larger than the reference value from DNS. This behaviour is well known (see e.g. \citet{osth2014need, Noack:2016jk}). 
The sparse Galerkin model obtained without the stability constraint (problem P1) also reproduces this behaviour. Conversely, the sparsified and stabilised model obtained from problem P2 has realistic temporal behaviour and resolves correctly the average energy and its fluctuation observed in DNS. This suggests that including in the optimisation information on how modal structures are supposed to interact with each other to satisfy the overall power budget is key to obtain sparse Galerkin models with adequate predictive ability. Note that the long-term performance of the models obtained in the present work is generally superior to that of models sparsified \emph{a posteriori}, using a LASSO-based approach \citep{rubini2020l1,RUBINI2020}. More specifically, LASSO-based models have been found to be temporally accurate over a time span comparable to that of the data used for the sparsification. Conversely, the present \emph{a priori} sparsified systems are temporally stable for much longer time horizons. In addition, numerical solutions converge to the asymptotic attractor for much larger perturbations of the initial conditions. We argue that this is likely the consequence of enforcing a stronger consistency between the modal structures and the corresponding Galerkin model. 

\begin{figure}\label{fig:Nijk_sparse}
  \centerline{\includegraphics[width=1\textwidth]{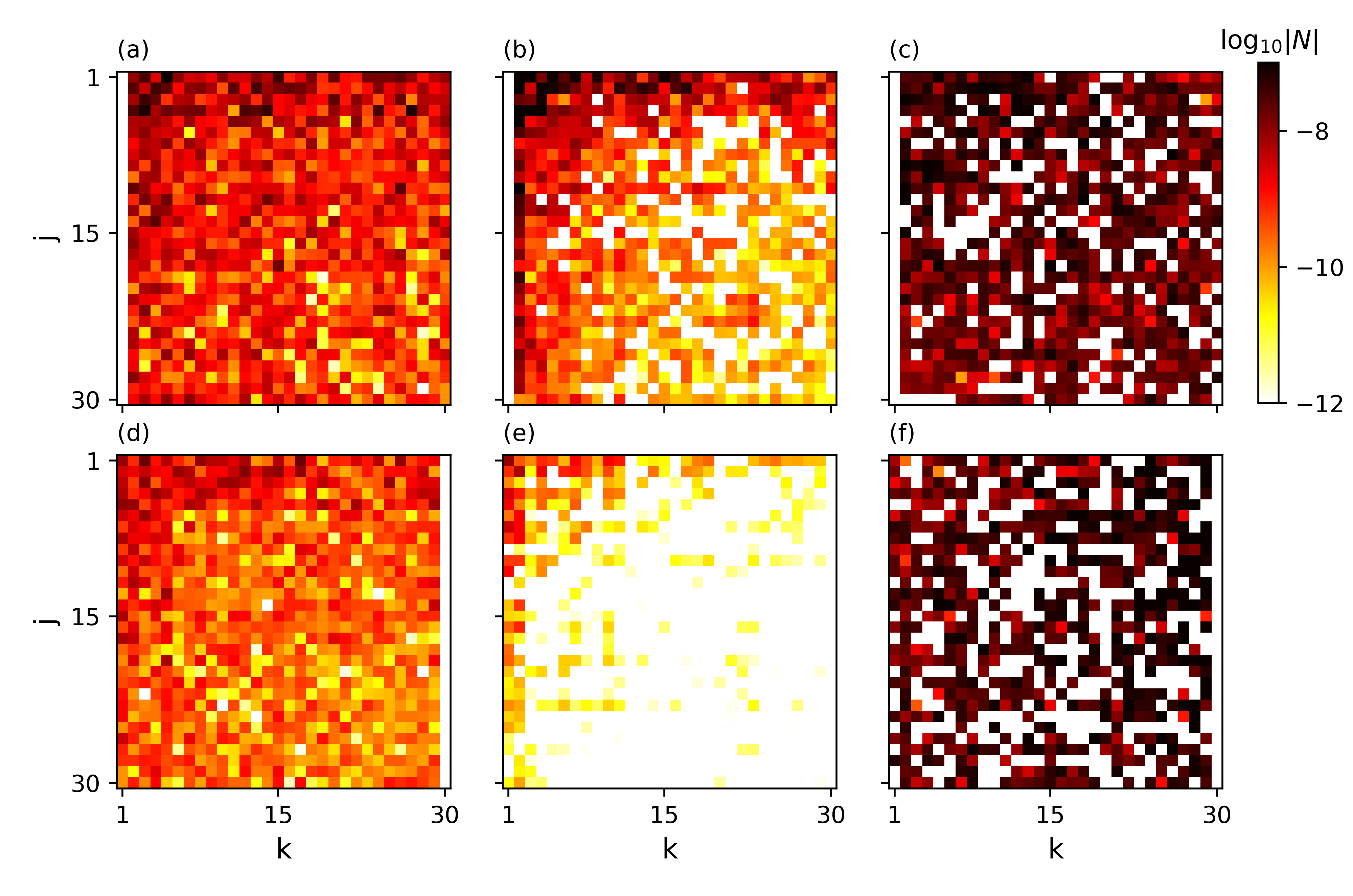}}
  \caption{Intensity of the average nonlinear energy transfer rate $\tilde{\mathsfbi{N}}$ for mode $i=1$ in models obtained from projection and formulations P1 and P2 in panels (a), (b) and (c), respectively. Panels (d), (e) and (f) show the same quantity for mode $i=30$.}
\end{figure}

The structure of the nonlinear energy transfer rate tensor $\tilde{\mathsfbi{N}}$ for the rotated Galerkin models is reported in figure \ref{fig:Nijk_sparse}, showing two slices for $i=1,30$ (first and second row, respectively). Panels (a) and (d) refer to the model obtained from Galerkin projection onto the original POD subspace and will be used as reference. Here the temporal coefficients are the projections onto the DNS snapshots. Data for the model obtained from problem P2 is shown in panels (b) and (e) while data for the model obtained with the P1 formulation is shown in panels (c) and (f). For these models, temporal coefficients are obtained from forward integration of the Galerkin models.
It can be observed that the model from formulation P2 displays a pattern of interactions consistent with the DNS data projected onto the original POD modes, both in terms of organisation across modes as well as in strength. The organisation of the interactions is similar to that in the dense Galerkin model obtained using the \emph{a-priori} stabilisation method of \citep{balajewicz2013low}, although a more aggressive reduction in the strength of the interactions between the high-index modes is observed. By contrast, energy interactions in the model obtained from the formulation P1 are orders of magnitude more intense, across all triads. This is a manifestation of the lack of dissipation and the consequential over-prediction of energy, across all modes. Interestingly, it can be observed that the $HL/LH$ asymmetry observed in figure \ref{fig:Energy_Analsis} is preserved throughout the sparsification. Conversely, this physical feature is lost in the \emph{a posteriori} approach  \citet{rubini2020l1} for numerical reasons. It is arguable that the \emph{a priori} approach is preferable for systems with a complex structure of $\mathsfbi{N}$. 

\begin{figure}
  \centerline{\includegraphics[width=1.1\textwidth]{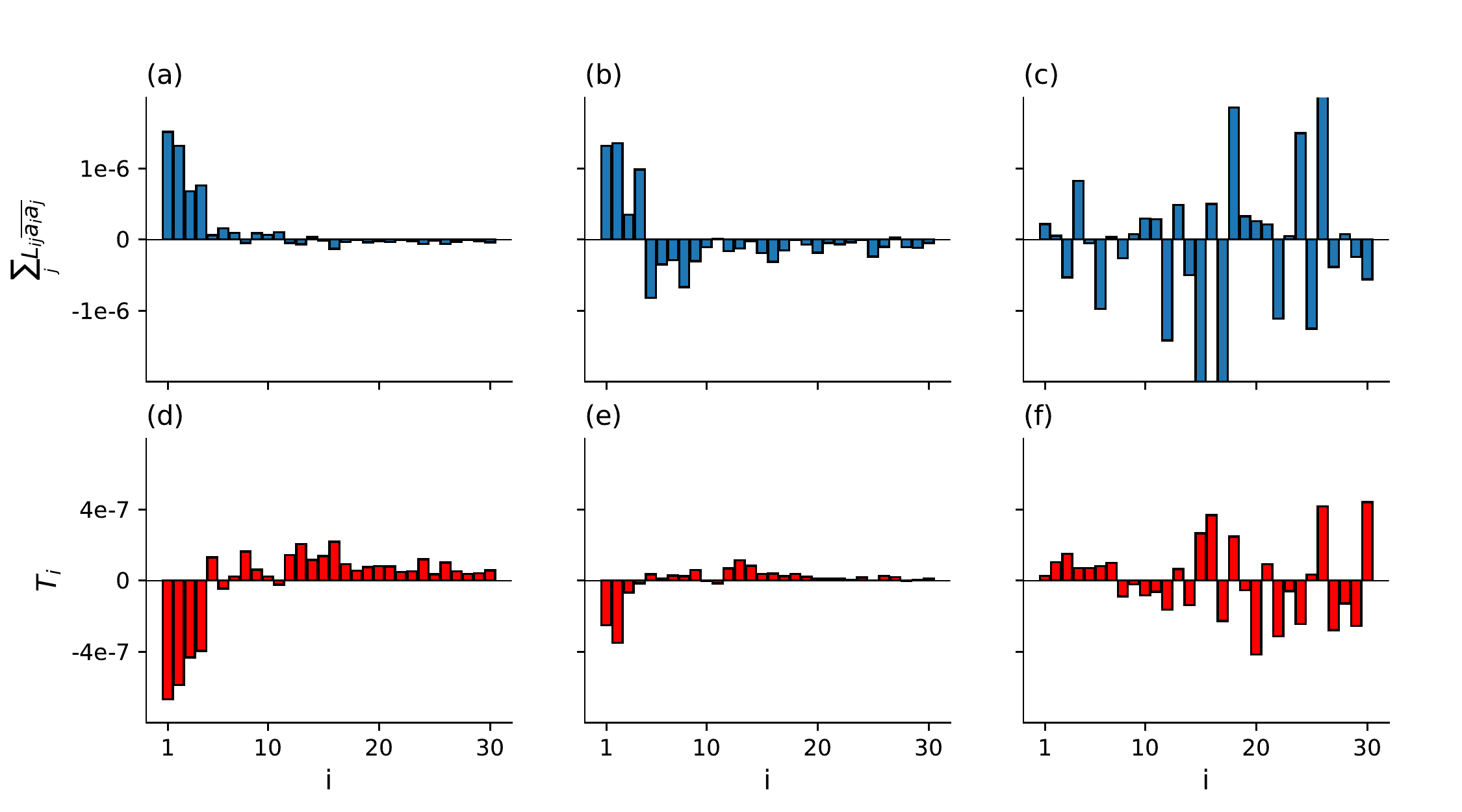}}
  \caption{Distribution of the average linear transfers $\sum_{j}\mathsf{L}_{ij}a_ia_j$ as a function of $i$ for the original POD-Galerkin model, and the models obtained from formulations P2 and P1, in panel (a), (b) and (c), respectively. Average nonlinear transfer rates $\mathsf{T}_i$ in panels (d), (e) and (f), for the same models. Temporal coefficients for this analysis are obtained from projection of the DNS data onto the original and rotated modes.} 
  \label{fig:Mean_Energy_balance}
\end{figure}

 
To conclude this section, we analyse the energy budget of the original and rotated Galerkin models, examining the linear and quadratic energy transfer terms in the average energy budget equation \eqref{eq:energy_avg} on a mode-by-mode basis, \citep{noack2008finite}. To characterise the total nonlinear energy transfer, we use the quantity
\begin{equation}\label{eq:T_i}
    \mathsf{T}_i = \sum_{j=1}^N \sum_{k=1}^N \mathsf{Q}_{ijk}\overline{a_ia_ja_k},\quad i = 1, \ldots, N
\end{equation}
satisfying $\sum_i \mathsf{T}_i = 0$ when the quadratic term conserves energy.
In figure \ref{fig:Mean_Energy_balance} we report data for three Galerkin models: the original POD-Galerkin model (first column) with temporal modes obtained from projection on the DNS data, the model obtained from the complete formulation P2 (second column) and the model obtained from formulation P1, with temporal coefficients obtained from temporal integration. All models have $N=30$ and, where applicable, $M/N=3$. 
%
The linear transfer term for the POD-Galerkin model exhibits a high production of kinetic energy in the first four modes, corresponding to the dominant structure and flow oscillation in the cavity, but only moderate dissipation in the remaining modes. This results into an inbalance between production and dissipation, leading to the over-prediction of kinetic energy as shown in figure \ref{fig:PDF_TKE}-(a). The mean nonlinear transfer rate $\mathsf{T}_i$ shown in panel (d), instead, displays the correct transfer direction, with energy injected from the large scales (negative transfer rate for low index modes) to the small scales (positive rate for the remaining modes). The model obtained from the solution of the complete formulation P2 displays a better balance between production and model dissipation, as larger dissipation in the high-index modes is observed, due to effect of the stability constraint \eqref{eq:optProb_2b}.  The distribution and the direction of the nonlinear transfers is also well preserved, although lower transfer rates are observed for all modes. This results from the combined effect of the sparsity promoting constraint, which reduces the magnitude of of the quadratic coefficients, and the reduction of the modal energies, due to the loss of optimality from the POD.  On the other hand, energy transfers in the model obtained from formulation P1, which does not include any dynamical information on the fluctuation kinetic energy budget \eqref{eq:energy_avg}, have a significantly different structure. Linear transfers show a markedly increased production and dissipation, while the organisation of nonlinear transfers across the hierarchy of modes has lost the original character.

These results show that to obtain a sparse model that is consistent with the flow physics observed in direct simulation it is of paramount importance to retain in the sparsification algorithm some information regarding the temporal dynamics. This is consistent with what observed by \citet{loiseau2018constrained}, who showed that an additional constraint on the energy conservation is needed to enhance the temporal accuracy of fluid model reconstructed from data. 


\section{Conclusions}
Scale interactions in turbulent flows are sparse, and the motion at any given length scale depends most prominently on the dynamics of a subset of all other scales. In addition to the recent work of \citet{schmidt2020bispectral}, this paper is one attempt at developing a model order reduction technique that leverages this fact, so that the analysis and interpretation of scale interactions in complex flows is facilitated. Here, we have proposed an \textit{a priori} sparsification approach, whereby a set of basis functions describing coherent structures that interact minimally with one another is sought for. Scale interaction sparsity is then defined by the sparsity of the quadratic coefficient tensor of the Galerkin model constructed from projection on this basis. As opposed to our previous sparsification approach \citep{rubini2020l1}, where model coefficients corresponding to weak nonlinear interactions between a pre-determined set of basis functions are pruned  \emph{a-posteriori} using $l_1$ regression, the present methodology maintains the exact link between the (sparse) Galerkin model and the modal structures utilised for the projection.

In practice, a set of $N$ basis functions is expressed as a rotation of $N$ Proper Orthogonal Decomposition (POD) modes within a larger POD subspace (of dimension $M > N$), with the idea of altering energy paths across the model by rotating the subspace utilised for projection whilst minimising the loss energy representation ability of the original POD basis. The rotation matrix is then found from the solution of a constrained optimisation problem where a) the energy loss with respect to the original POD basis is minimised and b) an inequality constraint involving the $l_1$ norm of the quadratic coefficient tensor associated to the rotated basis promotes sparsity. This formulation is augmented with a further constraint that enforces long-term temporal stability, following the approach of \citet{balajewicz2016minimal}. This optimisation problem depends on a single user-controllable parameter controlling the $l_1$ norm of the rotated quadratic coefficient tensor and the trade-off between energy reconstruction and sparsity.


To demonstrate the approach, we considered two-dimensional lid-driven cavity flow at Reynolds number $Re = 2 \times 10^4$, where the motion is chaotic and energy interactions are scattered in modal space. 
Results show that the energy reconstruction--sparsity trade-off depends on the model size $N$ and the ratio $M/N$. In particular, larger models can be sparsified more aggressively with less impact on the energy reconstruction ability. On the other hand, higher sparsity can be obtained by rotating the same $N$ POD basis functions within larger subspaces (a higher ratio $M/N$), since the increased flexibility allows shrinking to zero a larger fraction of quadratic interaction coefficients.  A deeper analysis of energy paths shows that the distribution of inter-modal energy transfers in the rotated model is qualitatively similar to that of the original POD-Galerkin system. More specifically, coefficients of the quadratic interaction tensor corresponding to large-scale/large-scale and the large-scale/small-scale interactions are preserved, but those defining weaker small-scale/small-scale interactions are shrunk to zero during the optimisation. This result is in agreement with the established picture of triadic interactions in two-dimensional flows and with previous results of sparsification of reduced order models \citep{rubini2020l1}. More interestingly, we observed that a physically-consistent organisation of the interactions and a stable long-term behaviour can only be obtained by augmenting the sparsification procedure with the temporal stability constraint. In fact, models obtained without such constraint inherit well-known temporal stability issues displayed by dense POD-Galerkin models.
%
%
%

Some aspects deserve further discussion. Unlike in $l_1$-regression-based sparsification methods \citep{loiseau2018constrained, rubini2020l1}, where the model coefficients are the optimisation variables, here these coefficients are cubic polynomial functions of the optimisation variables, the entries of the rotation matrix. The first consequence is that the optimisation problem is non-convex and many local optima, i.e. many different sets of basis functions, exist. It was observed that initial guesses close to the original POD basis, i.e. with good energy representation ability, repeatedly converge to the same optimal solution, which has a consistent physical interpretation. However, the lack of strong uniqueness guarantees (as for many other modal decomposition techniques) may render the approach questionable. Second, it is not possible to shrink to zero an arbitrary number of model coefficients, as the strong link between the basis functions and the Galerkin model must always be maintained. As a result, the present \emph{a-priori} sparsification technique produces relatively denser Galerkin models than the \textit{a posteriori} LASSO-based approach considered in \citet{rubini2020l1} for the same test case. Rotating the original POD basis into higher-dimensional subspaces, with ratio $M/N$ higher than that considered in the present work, to further increase sparsity is possible, although it would necessarily result in increased computational costs for the optimisation.

\vspace{0.2cm}

The authors gratefully acknowledge support for this work from the Air Force Office of Scientific Research (Grant No. FA9550-17-1-0324, Program Manager Dr D. Smith).

\appendix

\section{Sparsity-promoting effect of the $l_1$-based constraint}
\label{app:norma_l1}
In this appendix we visualise how the $l_1$ norm appearing in the constraint (\ref{eq:optProb_2b}) favours sparse solutions. For simplicity, we use a small ($N = 3$, $M = 5$) reduced order model of the lid-driven cavity flow considered in the previous sections. A more formal discussion based on the proximity operator theory can also be formulated, using the classical LASSO formulation as a starting point \citep{stat_learning}. 
\begin{figure}
  \centerline{\includegraphics[width=1.1\textwidth]{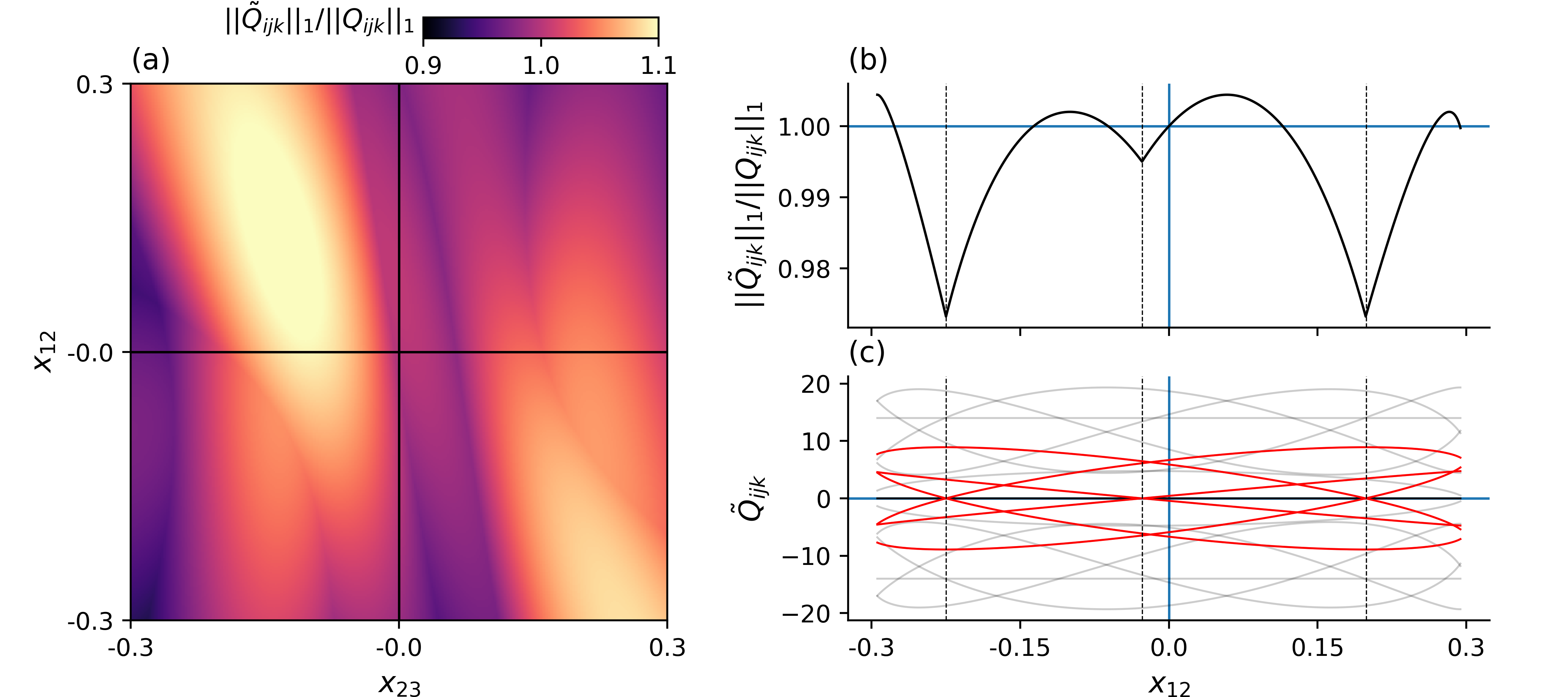}}
  \caption{Map of $||\mathsfbi{\tilde{Q}}||_1/||\mathsfbi{Q}||_1$ projected on the $\mathsf{X}_{12} - \mathsf{X}_{23}$ plane, panel (a). One dimensional cut of the same quantity along the coordinate $\mathsf{X}_{12}$, panel (b). The twenty-seven entries of the rotated $\tilde{\mathsfbi{Q}}$ as a function of $\mathsfbi{X}_{12}$. The six entries that vanish identically at one of the sharp points of  $||\mathsfbi{\tilde{Q}}||_1$ are highlighted in red.}
  \label{fig:Optimisation_constraint}
\end{figure}
In figure \ref{fig:Optimisation_constraint}-(a) the ratio $||\mathsfbi{\tilde{Q}}||_1/||\mathsfbi{Q}||_1$ is shown as a function of the entries $\mathsf{X}_{23}$ and $\mathsf{X}_{12}$ of the rotation matrix. When these two parameters are varied, all the other off-diagonal entries of $\mathsfbi{X}$ are set to zero. Several sharp valleys can be observed, arising from the non-smooth nature of the $l_1$ norm. Note that the $||\mathsfbi{\tilde{Q}}||_1/||\mathsfbi{Q}||_1$ is a non-convex cubic function of the entries of the rotation matrix and several local minima can be identified. Panel (b) shows a cut of panel (a) along the $\mathsf{X}_{12}$ coordinate, where minima are identified by the dashed vertical lines. Each minimum corresponds to a point where one of the entries of $\tilde{\mathsf{Q}}_{ijk}$ crosses the zero axis, as illustrated in panel (c). During the the optimisation the sparsity-promoting constraint pushes the solution towards one of these valleys, due to the strong gradient at the point of non-differentiability, resulting in a sparse coefficient tensor $\tilde{\mathsfbi{Q}}$. However, unlike in the \emph{a-posteriori} LASSO-based sparsification methods \citet{brunton2016discovering, rubini2020l1}, not all quadratic coefficients can be simultaneously set to zero by an arbitrary rotation, as clear from panel (c). In fact, the tensor $\mathsfbi{\tilde{Q}}$ depends nonlinearly on the rotation $\mathsfbi{X}$, while in LASSO-based methods the tensor coefficients are directly the optimisation variables of the problem.

\section{Dependence of the solution from the initial condition}\label{app:uniqueness}
Problem \eqref{eq:optimisation_problem_P2} is non-convex due to the constraint on the $l_1$ norm of the rotated quadratic interaction coefficient tensor $\tilde{\mathsfbi{Q}}$.
%
%
To understand how this feature affects the results of section \ref{sec:results}, we consider in this appendix different sets of initial guesses constructed as increasingly larger perturbations of the identity as
\begin{equation}\label{eq:random_X0}
    \mathsfbi{X} = \begin{bmatrix}\mathsfbi{I}_{N\times N} \\ \mathsfbi{0}  \end{bmatrix} + \epsilon \mathsfbi{R},
\end{equation}
with $\mathsfbi{R} \in \Re^{M \times N}$ a randomly generated rotation matrix, satisfying $\mathsfbi{R}^\top \mathsfbi{R} = \mathsfbi{I}_{N\times N}$. We generated ten random rotation matrices for five values of $\epsilon$ in the range $[10^{-8}, 10]$ and solved problem \eqref{eq:optimisation_problem_P2} by keeping $\xi = 2$ to target an intermediate density in the feasibility region, using $N = 30$ and $M = 90$.

\begin{figure}
  \centerline{\includegraphics[width=1\textwidth]{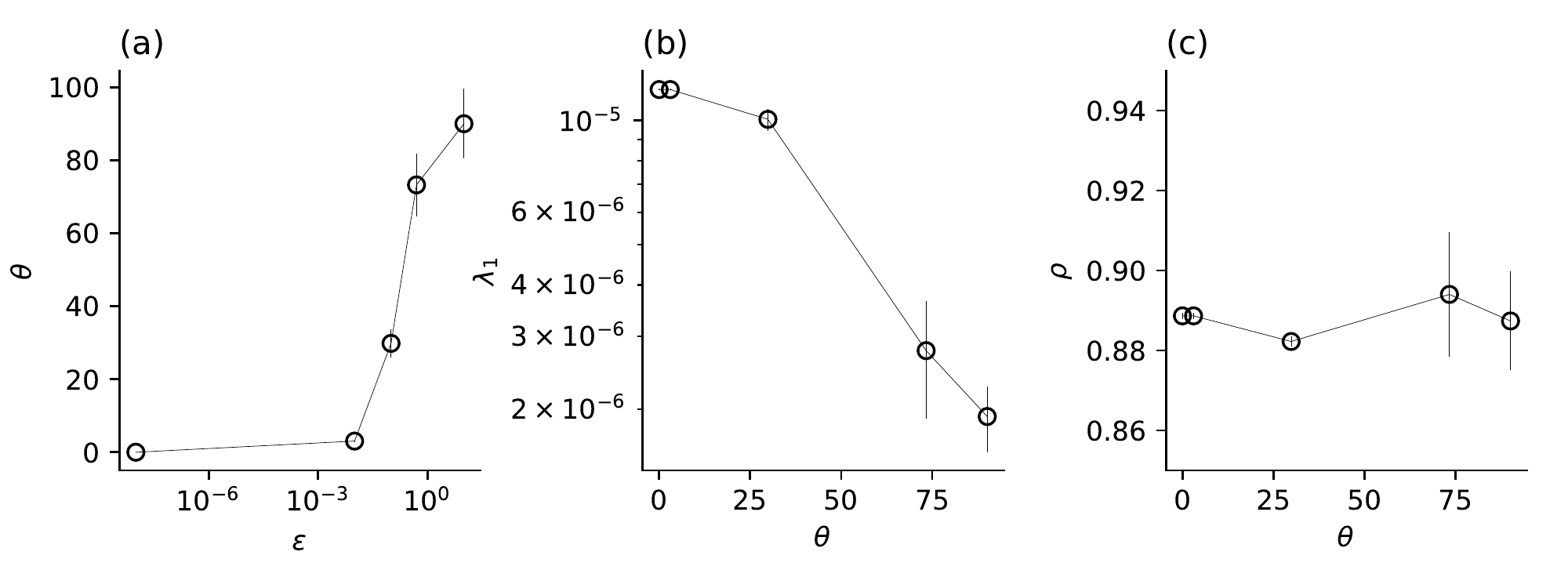}}
  \caption{Effect of the initial guess on the optimisation results. Panel (a): relation between the initial guess perturbation size $\epsilon$ and the average rotation angle $\theta$ away from the original POD basis. In panels (b) and (c): mean and standard deviation of the first modal energy $\tilde{\lambda}_1$ and density $\rho$, respectively, from rotated systems obtained from initial guesses of increasing distance from the original POD basis.}
  \label{fig:Uniqueness_of_the_solution} 
\end{figure}

Results are shown Figure \ref{fig:Uniqueness_of_the_solution}. Panel a) quantifies the effects of $\epsilon$ on the average rotation angle $\theta$ between columns of the initial guess and those of the matrix $[\mathsfbi{I}_{N\times N}, \mathsfbi{0}]^\top$. Circles and vertical bars identify the average and standard deviation of this quantity across the ten different samples. In panels (b) and (c), the largest modal energy $\tilde{\lambda}_1$ of the rotated temporal coefficients and the density $\rho$ of the optimal Galerkin model, respectively, are shown as a function of the average $\theta$. Results show that for small perturbations of the original POD basis, all initial guesses converge to the same solution, since $\tilde{\lambda}_1$ and $\rho$ are the same (up to the tolerances set in the optimisation).  Conversely, initial guesses corresponding to larger (random) rotation of the original POD basis converge to different solutions, corresponding to modal structures with lower reconstruction ability, but similar model density.

\section{Optimisation of tensor operations and scaling of costs}\label{app:comp_cost}
The sparsity promoting constraint involves polynomial functions of the optimisation variables. An analytical expression of the gradient of this constraint with respect to the rotation $\mathsfbi{X}$ can be also obtained, enabling fast gradient-based optimisation to be utilised. In addition, a careful examination of the tensor operations involved in the computation of the constraint and its gradient shows that a significant reduction of the scaling of costs can be obtained by reorganising some tensorial computations and using intermediate temporary variables (see \citet{PhysRevE.90.033315}). For instance, a naive calculation of all entries of the rotated tensor \eqref{eq:rotation}, required in the evaluation of the sparsity-promoting constraint \eqref{eq:optProb_b}, takes $\mathcal{O}(M^3 N^3)$ operations since six different nested for loops (one for each index) are involved in total.
However, a careful examination of \eqref{eq:rotation} shows that the rotated tensor can also be obtained by first introducing the auxiliary temporary tensors $\mathsfbi{A} \in \Re^{M\times M \times N}$ and $\mathsfbi{B} \in \Re^{M\times N \times N}$, and then computing
\begin{subequations}\label{eq:rotation_optimised}
\begin{align}
    \mathsf{A}_{qri} &= \sum_{p=1}^M \mathsf{Q}_{pqr} \mathsf{X}_{pi}, \\
    \mathsf{B}_{rij} &= \sum_{r=1}^M \mathsf{A}_{qri} \mathsf{X}_{qj}, \\
    \tilde{\mathsf{Q}}_{ijk} &= \sum_{q=1}^M  \mathsf{B}_{rij} \mathsf{X}_{rk},
\end{align}
\end{subequations}
with $i, j, k = 1, \ldots, N$ and $q, r = 1, \ldots, M$. Collectively, these steps only take $\mathcal{O}(M^3 N) + \mathcal{O}(M^2 N^2) + \mathcal{O}(M N^3)$ operations, as can be inferred from the summation indices.

On the other hand, the gradient of the sparsity promoting constraint with respect to the entries $\mathsf{X}_{mn}$ of the rotation matrix $\mathsfbi{X}$ requires the computation of
\begin{equation}\label{eq:dl1dX}
        \frac{\partial|| \tilde{\mathsfbi{Q}}||_1}{\partial \mathsf{X}_{mn}} = \sum_{i, j, k =1}^N\frac{\partial \tilde{\mathsf{Q}}_{ijk} }{\partial \mathsf{X}_{mn}} \text{sign}(\tilde{\mathsf{Q}}_{ijk}),
\end{equation}
where the gradient $\partial \tilde{\mathsf{Q}}_{ijk}  / \partial \mathsf{X}_{mn}$ can be expanded using the Kronecker's delta $\delta_{ij}$ as
\begin{equation}\label{eq:dQdX}
       \frac{\partial \mathsf{\tilde{Q}}_{ijk}}{\partial \mathsf{X}_{mn}}  = \sum_{p, q, r = 1}^{M}\mathsf{Q}_{pqr}(\mathsf{X}_{pi}\mathsf{X}_{qj} \delta_{kn} \delta_{rm}     + \mathsf{X}_{pi}\mathsf{X}_{rk} \delta_{jn} \delta_{qm} + \mathsf{X}_{rk}\mathsf{X}_{qj} \delta_{in} \delta_{pm}).
\end{equation}
Overall, a naive computation of \eqref{eq:dQdX} would require $\mathcal{O}(M^4N^4)$ operation, since the tensor expressions involves a total of eight indices. 
A similar reorganisation of the operations involved in \eqref{eq:dQdX} as described above leads to tensor operations over four tensor indices only, with computational costs scaling  similarly to the rotation.

Lastly, we underline that in the present implementation we chose to store the whole tensors \eqref{eq:rotation}-\eqref{eq:dQdX} and \eqref{eq:dl1dX} in memory. This approach is computational efficient since enable to use optimised packages for tensorial calculus but it is very expensive memory-wise. A further optimisation can be obtained by assembling \eqref{eq:rotation} and \eqref{eq:dl1dX} without fully storing of the intermediate steps.

\bibliographystyle{jfm}
\bibliography{jfm-instructions}

\end{document}